%% file: main.tex
\documentclass{article}

\usepackage[main, final]{neurips_2026}

\usepackage[utf8]{inputenc} 
\usepackage[T1]{fontenc}    
\usepackage{hyperref}       
\usepackage{url}            
\usepackage{booktabs}       
\usepackage{amsfonts}       
\usepackage{nicefrac}       
\usepackage{microtype}      
\usepackage{xcolor}         
\usepackage{amsmath}
\usepackage{graphicx}
\usepackage{wrapfig}
\usepackage{enumitem}
\usepackage{booktabs}
\usepackage{makecell}
\newcommand{\pgstd}[1]{{\scriptsize\,$\pm$#1}}
\newcommand{\pgbest}[1]{\textbf{#1}}
\newcommand{\pgbudget}[1]{{\scriptsize\itshape #1}}
\title{Learning When Visual Context Matters for Mouse Behavior Analysis}

\author{%
  Weihan Li$^{1}$ \quad Jingyang Ke$^{1}$ \quad Qinheng Pu$^{2}$ \quad Yule Wang$^{1}$ \quad Chengrui Li$^{1}$ \quad Anqi Wu$^{1}$ \\
  $^{1}$School of Computational Science and Engineering, Georgia Tech \\
  $^{2}$Wallace H.~Coulter Department of Biomedical Engineering, \\
  Georgia Tech and Emory University
}

\begin{document}

\maketitle

\begin{abstract}
Understanding mouse behavior at scale is a foundational problem in neuroscience, with two core tasks: classifying frames into known behaviors and segmenting recordings into recurring patterns without labels. Pose-based methods are the standard approach but face fundamental limitations: pose coordinates capture only skeletal information, discard environmental context, and degrade under frequent occlusions in social scenarios. Mouse videos can provide the visual context that pose lacks, yet not all frames benefit equally, and adding visual features uniformly to every frame is suboptimal. We find that a pose-based motion encoder, trained with auxiliary supervision from videos, can identify where it struggles: with a discrete representation, the encoder learns a codebook of recurring patterns, and frames whose motion does not fit this codebook produce large quantization residuals. We show that these high-residual frames are where visual context tends to help most: residual-guided frame selection consistently outperforms uniform selection across budgets, and matches or exceeds dense visual processing in most settings. We therefore propose ResiFuse, a framework that pretrains a motion encoder on pose with an auxiliary cross-modal loss and integrates vision context only on high-residual frames at fine-tuning time. We evaluate on three mouse behavior datasets and show improvements over pose- and video-based baselines on most tasks, with the largest gains on the most socially complex dataset. Code is available at \url{https://github.com/BRAINML-GT/ResiFuse}.
\end{abstract}

\section{Introduction}
\label{sec:intro}
 

Understanding mouse behavior is a central problem in neuroscience, requiring the conversion of long video recordings into interpretable behavioral structure. Researchers typically approach this through two complementary tasks: \emph{supervised classification}, which assigns each frame to predefined behavior labels such as running, grooming, chasing, or huddling, enabling annotation of new videos; and \emph{unsupervised segmentation}, which discovers recurring patterns not previously identified, revealing novel motifs beyond known behaviors.

To address these tasks, after extracting keypoints from video with pose estimation tools~\citep{mathis2018deeplabcut, pereira2022sleap, biderman2024lightning}, a common pipeline in modern machine learning and computational neuroscience adopts a two-stage framework for behavior analysis. First, a model is pretrained on pose keypoints and/or videos using self-supervised learning to learn useful representations. The pretrained model is then fine-tuned on the same modalities with task-specific objectives, such as classification or segmentation. This paradigm of representation learning followed by downstream fine-tuning has become standard in mouse behavior modeling~\citep{azabou2023relax, stoffl2024elucidating}.

 
Pose-based approaches~\citep{hsu2021b, luxem2022identifying, weinreb2024keypoint, azabou2023relax,stoffl2024elucidating} have driven significant progress, but they face three fundamental challenges. First, pose only captures skeletal coordinates and discards finer visual details such as body texture, contact appearance, and the motion of small body parts that may distinguish behaviors with similar coarse kinematics. Second, pose discards environmental context such as scene structure, lighting, and surrounding objects, which may be needed to disambiguate behaviors with similar pose patterns. Third, pose estimation is sensitive to occlusions: in social scenarios, mice frequently occlude each other, causing pose estimates to fluctuate dramatically and lose information precisely at the moments when behavior is most informative.
 
Raw video contains the visual context that pose lacks, and prior work has shown that integrating video features can help~\citep{sun2024video, wang2025self,liu2025castle}.
However, the failures above do not occur on every frame: pose is sufficient on some frames, and a subset of frames truly need additional visual context.
Adding visual features uniformly across all frames is therefore both computationally expensive and suboptimal.
The key challenge is not simply combining pose and video, but knowing \emph{where} visual context is needed.

\begin{wrapfigure}{r}{0.5\textwidth}
\vspace{-10pt}
\centering
\includegraphics[width=0.5\textwidth]{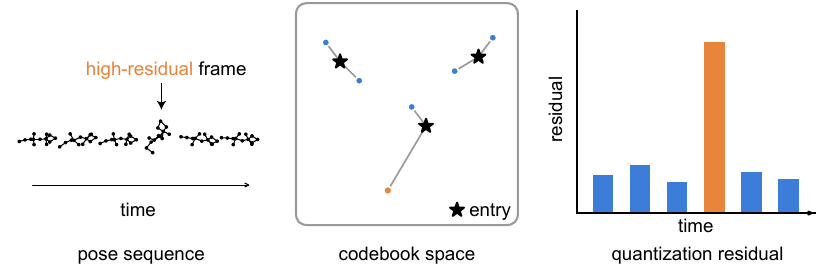}
\caption{\textbf{The quantization residual.} The motion encoder maps each frame to the nearest entry in a codebook of typical motion patterns. Most frames are well-represented; some frames have motion patterns that lie far from any codebook entry, producing a large quantization residual (orange).}
\label{fig:concept}
\vspace{-10pt}
\end{wrapfigure}

We tackle this challenge by following the two-stage pipeline and using both video and pose, but our goal is not simply to combine video and pose, which may be suboptimal, but to determine where and how visual information should be incorporated. We propose a two-stage learning paradigm named ResiFuse. Specifically, in the pretraining stage, we first learn a \emph{motion encoder}, a pose-based network trained with auxiliary supervision from video that compresses pose sequences into a compact latent representation. A key insight is that the motion encoder itself shows where it struggles. When trained with a discrete representation~\citep{van2017neural}, the encoder learns a finite codebook of recurring patterns, mapping each frame to its nearest entry (Figure~\ref{fig:concept}). While most frames are well represented, some are not, leading to a large distance to the nearest entry, which we term the \emph{quantization residual}. For example, during sudden movements, the encoder may fail to represent the frame compactly, resulting in a large residual. 
At fine-tuning stage, the residual guides where visual processing is applied: visual features are extracted only for high-residual frames and fused with the motion representation. Across a range of vision encoders, this residual-guided selection consistently outperforms alternatives such as uniform sampling, indicating that high-residual frames are precisely where visual context is most beneficial.

A natural question is whether simply processing every frame with a vision encoder would be simpler and better.
We find that it usually is not: \emph{residual-guided selection matches or exceeds dense visual processing in most settings}.
Crucially, the selection signal is parameter-free and interpretable: it is a natural by-product of motion encoder training, with no additional learning required, and each selected frame can be traced back to a specific frame where the motion representation is least reliable. While reconstruction error in input space has been used as an anomaly signal~\citep{gong2019memorizing, park2020learning, biderman2024lightning}, the quantization residual is a distinct quantity computed directly in the encoder's latent space, and to our knowledge, it has not previously been used as a selection signal for pose-video fusion.

 
We evaluate ResiFuse on three mouse behavior datasets of increasing social complexity:
Open Field~\citep{sturman2020deep}, where one mouse explores an open field;
CalMS21~\citep{sun2021multi}, which captures pair interactions between a resident and an intruder mouse;
and Mouse Triplet~\citep{sun2023mabe22}, which records social interactions among three mice.
Across these datasets, pose quality and the prevalence of occlusion vary substantially, providing a test of how the relationship between quantization residuals and visual value generalizes. On these datasets, ResiFuse improves over pose- and vision-based baselines on four of six tasks, with the largest gains on the most socially complex dataset, and matches or exceeds its own dense-vision variant at partial frame budgets in most settings.

 
\paragraph{Contributions.} First, we propose ResiFuse, a framework that integrates pose and video through selective fusion: visual features are extracted only on a subset of frames identified by a parameter-free signal, and the same architecture supports both supervised classification and unsupervised segmentation.
Second, we show that the selection signal can be the quantization residual of a pretrained motion encoder, a free by-product of training, and to our knowledge has not previously been used for this purpose.
Third, we conduct experiments on three mouse behavior datasets spanning single-animal, pair, and group settings, analyzing how ResiFuse's effectiveness varies across selection strategies, frame budgets, vision encoders, and pretraining choices.

\begin{figure*}
\vspace{-10pt}
    \centering
    \includegraphics[width=1\linewidth]{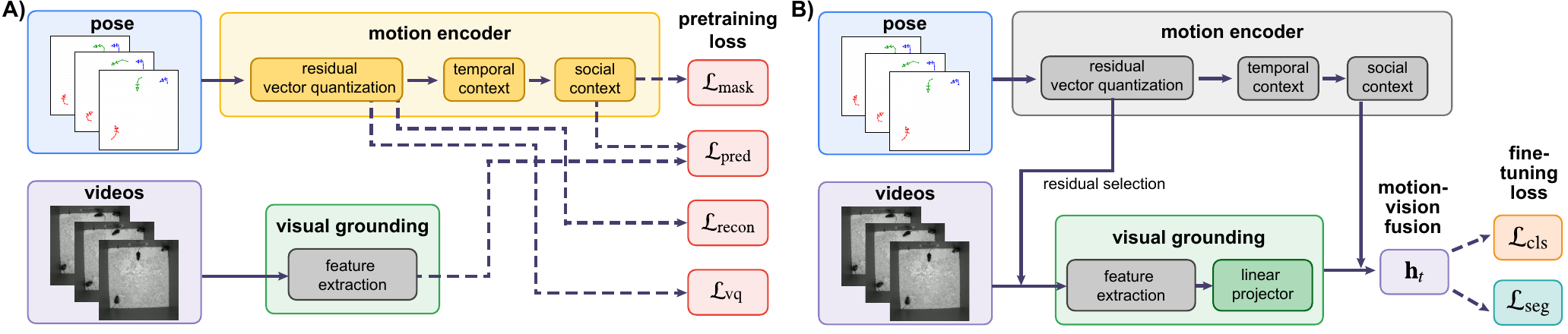}
   \caption{\textbf{ResiFuse framework.} 
(\textbf{A}) \textit{Pretraining}: a motion encoder compresses pose sequences through residual vector quantization (RVQ), temporal context, and social context modules, trained with four objectives ($\mathcal{L}_\text{mask}$, $\mathcal{L}_\text{pred}$, $\mathcal{L}_\text{recon}$, $\mathcal{L}_\text{vq}$). The cross-modal prediction loss $\mathcal{L}_\text{pred}$ uses visual features from every frame as auxiliary supervision, shaping the residual to indicate when pose alone is insufficient. 
(\textbf{B}) \textit{Fine-tuning}: the motion encoder is frozen. Its quantization residual selects the frames where visual features are extracted; the resulting motion-vision fused representation $\mathbf{h}_t$ supports both supervised classification and unsupervised segmentation as downstream tasks.}
    \label{fig:arch}
    \vspace{-10pt}
\end{figure*}

\section{Method}
 
 
ResiFuse is built around a simple principle: identify the frames where pose alone is insufficient, and concentrate visual features on those frames at fine-tuning.
The framework has two stages (Figure~\ref{fig:arch}).
In \textbf{pretraining}, a motion encoder is trained on pose in a self-supervised manner. An auxiliary cross-modal loss uses video features as additional supervision, further enhancing the residual's effectiveness as a selection signal.
In \textbf{fine-tuning}, the motion encoder is frozen, and we use its quantization residual to identify the frames where visual context is most needed, integrating visual features only on those frames.
The selection signal is parameter-free and interpretable, and the same architecture supports both supervised classification and unsupervised behavior segmentation.
 
\subsection{Pretraining the Motion Encoder}
\label{sec:method-pretrain}
 
The goal of pretraining is to learn a motion representation that captures as much behavioral structure as possible from pose alone, while producing a residual signal that reflects where it falls short.

\paragraph{Input features.} For each animal over $L$ frames, we concatenate joint coordinates, normalized by frame size, with their confidence scores, and, in multi-animal datasets, the relative position of the nearest other animals (Appendix~\ref{app:pretrain-details}). A temporal tokenizer maps these features to a continuous latent representation $z_e \in \mathbb{R}^{L \times D}$ ($D$ is the latent dimension), with one latent vector per frame that reflects temporal context rather than a single static pose (Appendix~\ref{app:implementation}).

\paragraph{Residual vector quantization.}
We discretize $z_e$ using Residual Vector Quantization (RVQ)~\citep{zhang2023generating, guo2024momask}, which replaces continuous latent with a sum of two codebook entries:
\begin{equation}
z_q^{(1)} = \text{VQ}_1(z_e), \quad z_q^{(2)} = \text{VQ}_2(z_e - z_q^{(1)}), \quad z_q = z_q^{(1)} + z_q^{(2)},
\label{eq:rvq}
\end{equation} where $\text{VQ}_i$ maps each vector to its nearest entry in codebook $\mathcal{C}_i$. Codebooks are updated by exponential moving averages (Appendix~\ref{app:implementation}).
Here $z_q^{(1)}$ captures a coarse, typical motion pattern from the first codebook, and $z_q^{(2)}$ adds a finer correction from the second; their sum $z_q$ is the quantized approximation of $z_e$.
Critically, quantization leaves a residual:
\begin{equation}
r_t = \| z_{e,t} - z_{q,t} \|_2,
\label{eq:residual}
\end{equation} which measures how well the codebooks can represent the motion at frame $t$, and which we use in the fine-tuning stage as a signal of representation quality. This residual is related to but mechanistically different from input-space reconstruction error in other domains~\citep{gong2019memorizing, park2020learning}: it is computed in latent space, measuring whether the encoder's continuous latent $z_{e,t}$ falls within the typical motion manifold spanned by the learned codebook, and directly captures the motion encoder's confidence in its own representation. Empirically, the two are only moderately correlated and diverge mainly on occluded frames (Appendix~\ref{app:residual-recon}).

\paragraph{Contextual aggregation.} \label{sec:contextual}
The quantized tokens $z_q$ are per-frame codes with only lightweight temporal mixing from the tokenizer, while behavior depends on richer temporal context and, in multi-animal recordings, the states of other animals. We capture these dependencies in two stages.

(1) \emph{Temporal context:} A self-attention Transformer attends across the time dimension for each animal independently, producing a temporally contextualized representation $e^{\text{tem}}_{a,t} = \text{Attn}_\text{tem}(z_{q,a,t}, \{z_{q,a,t'}\}_{t'=1}^{L})$ for animal $a$ at frame $t$.

(2) \emph{Social context:} For multi-animal recordings, a second Transformer applies cross-animal attention at each time step to capture interactions like chasing or huddling. The final output, $e_{\text{motion},a,t} = \text{Attn}_\text{soc}(e^{\text{tem}}_{a,t}, \{e^{\text{tem}}_{a',t}\}_{a' \in \mathcal{A}})$, relates each animal's pose dynamics to others in the set $\mathcal{A}$. 

This matters because spatial proximity alone is not enough to characterize an interaction: two mice may be close together while merely passing by each other, or while genuinely engaged in a chase or contact behavior. Distinguishing these cases requires reasoning about how each animal's pose dynamics relate to the others, which the cross-animal attention captures by comparing temporally contextualized representations across animals rather than by spatial distance. A post-hoc analysis confirms this empirically: attention is driven by interaction status rather than spatial proximity (Appendix~\ref{app:social_attention}). The resulting $e_{\text{motion},a,t}$ serves as the motion encoder's final output.

\paragraph{Training objectives.} \label{sec:method-objectives}
Pretraining combines four complementary objectives:
\begin{equation}
\mathcal{L}_{\text{pretrain}} = \mathcal{L}_{\text{mask}} + \lambda_{\text{recon}}\,\mathcal{L}_{\text{recon}} + \lambda_{\text{vq}}\,\mathcal{L}_{\text{vq}} + \lambda_{\text{pred}}\,\mathcal{L}_{\text{pred}}.
\label{eq:pretrain-loss}
\end{equation} \emph{(1) Masked pose reconstruction} ($\mathcal{L}_{\text{mask}}$): a subset of frames is masked, and the motion encoder predicts their normalized joint coordinates, encouraging it to capture fine-grained temporal dynamics.

\emph{(2) RVQ reconstruction} ($\mathcal{L}_{\text{recon}}$): the quantized representation $z_q$ are passed through a decoder and matched back to the input pose sequences, ensuring the discrete $z_q$ preserve enough information.
 
\emph{(3) RVQ commitment} ($\mathcal{L}_{\text{vq}}$): the loss that stabilizes codebook learning~\citep{van2017neural}.
 
\emph{(4) Cross-modal visual prediction} ($\mathcal{L}_{\text{pred}}$): the motion encoder predicts visual features from the frozen vision encoder at multiple future horizons using a cosine similarity loss. 

The first three objectives form a pose-based self-supervision: masked pose reconstruction and RVQ reconstruction encourage the encoder to capture pose structure, and RVQ naturally produces a residual that reflects how well each frame fits the learned codebook. We additionally introduce an auxiliary cross-modal objective that predicts the features of a frozen vision encoder from the motion representation. Under this objective, codebook entries must support not only pose reconstruction but also visual feature prediction. If a frame's visual content cannot be predicted from a typical codebook entry, the loss pulls that frame's latent toward a position that no typical entry sits near, producing a large residual. 

As a consequence, three types of frames produce a large residual: \emph{(i)} frames where pose alone is insufficient to characterize the moment, e.g., during sudden movements or at transitions between behaviors where similar poses map to different behaviors; \emph{(ii)} frames where the visual content cannot be inferred from pose alone, e.g., a moment of close social interaction where one mouse partially occludes another, leaving its pose unreliable while the visual content remains informative; and \emph{(iii)} frames with pose estimation errors, e.g., bedding mistaken for part of a mouse, which raise the residual when injected synthetically. Appendices~\ref{app:quantify}--\ref{app:pose-noise} analyze these cases. The cross-modal prediction objective uses visual features from every frame of the training videos, while the selection mechanism applies only at fine-tuning time. Full formulations, masking strategy, and loss weights are in Appendix~\ref{app:pretrain-details}.

\subsection{Residual-Guided Selective Fusion}
\label{sec:method-fusion}
 
The central component of ResiFuse is a fine-tuning stage that uses the quantization residual to decide where to apply visual processing.
While the cross-modal pretraining objective uses visual features from every frame, at fine-tuning time we find that selectively integrating vision on a subset of frames matches processing every frame in most settings and can exceed it (Section~\ref{sec:residual-budgets}).
 
\paragraph{Design space of selection signals.}
The challenge is to determine which frames need visual context at fine-tuning time. We organize this design space by the source of the selection signal at each frame, with three categories. \textbf{Pose-side signals} are derived directly from the pose tracker. A natural example is keypoint confidence, $r_t^{\text{conf}} = \frac{1}{K}\sum_k c_{t,k}$, where $c_{t,k}$ is the per-joint confidence from the pose estimator and $K$ is the total number of joints. Such signals capture tracking failures but provide no direct information about behavioral content. \textbf{Learnable selectors} pass the motion encoder's latent through a small network and train it end-to-end. A typical instantiation is $r_t^{\text{learn}} = \sigma_\theta(z_{e,t})$, an MLP optimized through the downstream task loss with hard top-$k$ selection via straight-through gradients~\citep{bengio2013estimating}. This approach adds extra parameters and ties the selector to a specific downstream task, requiring re-training when the task changes. \textbf{Quantization residual.} We propose using the quantization residual $r_t = \|z_{e,t} - z_{q,t}\|_2$, a free by-product of motion encoder pretraining that directly reflects the motion encoder's own representation quality. To our knowledge, the residual has not previously been used as a selection signal for pose-video fusion. It requires no additional parameters or training, can be precomputed offline once the motion encoder is pretrained, and is grounded in the pretraining objective. We compare these three alternatives empirically in Section~\ref{sec:ablations}.

\paragraph{Frame selection.}
For each pose sequence, we compute the residual $r_t$ from Eq.~\ref{eq:residual} at every frame and select the frames with the largest residuals: $\mathcal{S} = \text{Top-}k \big( \{r_t\}_{t=1}^{L} \big),\ k = \lfloor L \cdot \text{budget} \rfloor$, where $\text{budget} \in [0, 1]$ is the fraction of frames we process with the vision encoder.
$\mathcal{S}$ is the set of \emph{visually active} frames. By doing so,
each selected frame can be traced back to a specific frame where the motion representation is least reliable.

\paragraph{Visual feature integration.} \label{sec:ht}
Because $\mathcal{S}$ depends only on pose, it is known before any video frame is decoded, so only frames in $\mathcal{S}$ are decoded and passed to the vision encoder.
The visual features on selected frames are projected into the motion representation's dimension through a learned linear layer, while the visual input on non-selected frames is set to zero.
We fuse motion and visual features at the feature level: $h_t = \text{LN}_f \left( \text{LN}_m(e_{\text{motion},t}) + \mathbb{1}[t \in \mathcal{S}] \cdot \text{LN}_v(\text{proj}(v_t)) \right)$, where $\mathbb{1}$ is the selection mask and $\text{proj}$ is a learned linear layer. The three LayerNorms keep the fused feature at consistent magnitude across selected and non-selected frames. The fused $h_t$ is used for downstream tasks.

 
\subsection{Downstream Tasks}
\label{sec:method-downstream}
 
Both supervised classification and unsupervised segmentation build on the same fused representation $h_t$ and share the same frozen motion encoder; they differ only in how $h_t$ is processed after fusion.
During fine-tuning, all pretrained motion encoder components are frozen; only a small number of task-specific parameters are trained. Full details are in Appendix~\ref{app:sup-details}.
 
\paragraph{Supervised classification.}
Given per-frame behavior labels (e.g., grooming, chasing, huddling), the goal is to predict a behavior label for every frame in a test sequence.
A classifier takes the fused representation $h_t$ and predicts per-frame behavior labels (Appendix~\ref{app:implementation}).
 
\paragraph{Unsupervised segmentation.}
Without access to any behavior labels, the goal is to partition continuous recordings into clusters that correspond to distinct, recurring behaviors.
We follow the Deep Embedded Clustering (DEC)~\citep{xie2016unsupervised} to discover behavior clusters without labels, operating on the continuous fused representation $h_t$.
While the motion encoder already produces discrete RVQ codes, those codes are optimized to support the pretraining objectives (pose reconstruction, visual prediction) rather than to define behavior categories. Clustering on $h_t$ instead admits finer-grained behavioral structure and additionally incorporates visual information from selected frames. Empirically, we find $h_t$ clusters better than RVQ codes (Appendix~\ref{app:rvq_vs_dec}).
We intentionally use a larger number of clusters than the number of behavior classes, allowing the model to discover fine-grained patterns. At evaluation time, multiple clusters can be mapped to the same behavior class through a majority-vote mapping on a labeled training set.
To prevent representation collapse during clustering, we apply a variance-regularization term that penalizes dimensions with low variance.

\section{Experiment}
\label{sec:experiment}

\paragraph{Datasets.} We evaluate on three mouse behavior datasets of increasing social complexity.
\begin{itemize}[leftmargin=*,itemsep=0pt,topsep=2pt,parsep=0pt]
\item \textbf{Open Field}~\citep{sturman2020deep}: A single mouse explores an open arena, annotated with two individual behaviors (grooming, rearing; supported and unsupported rearing are merged, Appendix~\ref{app:experiment}). 20 top-down videos ($\sim$10\,min each at 25\,Hz), 13 keypoints per mouse.
\item \textbf{CalMS21}~\citep{sun2021multi}: A resident mouse interacts with an intruder, annotated with three social behaviors (attack, investigation, mount). 89 videos ($\sim$4.8\,min each at 30\,Hz), 7 keypoints per mouse.
\item \textbf{Mouse Triplet}~\citep{sun2023mabe22}: Three mice interact in an open arena, annotated with four social behaviors (chase, huddle, oral contact, oral-genital contact). 2{,}614 videos (60\,s each at 30\,Hz), 12 keypoints per mouse. We re-estimate pose with DeepLabCut~\citep{mathis2018deeplabcut} to reduce tracking artifacts, and all methods use the same re-estimated pose.
\end{itemize}

\paragraph{Baselines.} We compare ResiFuse (using VideoPrism~\citep{zhao2024videoprism} as the vision encoder) against the following baselines and report macro-averaged F1 over five seeds on the test set across behavior classes. Full setting details are in Appendix~\ref{app:experiment}.
\begin{itemize}[leftmargin=*,itemsep=0pt,topsep=2pt,parsep=0pt]
\item \textbf{BAMS}~\citep{azabou2023relax}: A self-supervised, pose-only method that integrates short-term and long-term temporal modeling through multi-scale mutual information maximization.
\item \textbf{hBehaveMAE}~\citep{stoffl2024elucidating}: A self-supervised, pose-only method based on hierarchical masked autoencoding, which learns representations across multiple temporal and spatial scales.
\item \textbf{Keypoint-MoSeq}~\citep{weinreb2024keypoint}: An unsupervised pose-only method that segments behavior into discrete syllables via autoregressive hidden Markov models. 
\item \textbf{BEAST}~\citep{wang2025self}: A self-supervised video method pretrained on raw videos via masked autoencoding and temporal contrastive learning, applied to all frames of the input video; supervised results use its native segmentation head.
\item \textbf{Pose coordinates / VideoPrism features / VideoPrism + pose coordinates}: normalized keypoint coordinates, per-frame VideoPrism features, or their concatenation, fed to the same classifier.
\item \textbf{Hand-crafted features}: dataset-specific pose features from \citet{sturman2020deep} with velocities (Open Field), MARS~\citep{segalin2021mouse} (CalMS21), and a MABe22 baseline~\citep{sun2023mabe22} (Mouse Triplet).
\end{itemize}


\subsection{Main Results}
\label{sec:main-results}

\begin{table*}[t]
\centering
\caption{Macro-averaged F1 (mean ± std) across three mouse-behavior benchmarks under \emph{supervised classification} and \emph{unsupervised segmentation}. Keypoint-MoSeq (KPMS) applies only to the unsupervised setting. For ResiFuse, we also report the best frame budget per cell, where the chosen budget is shown in the row immediately below ($r$=fraction of frames receiving vision).}
\label{tab:main}
\resizebox{\textwidth}{!}{%
\begin{tabular}{l ccc ccc}
\toprule
& \multicolumn{3}{c}{\textbf{Supervised}} & \multicolumn{3}{c}{\textbf{Unsupervised}} \\
\cmidrule(lr){2-4} \cmidrule(lr){5-7}
Method
& \makecell{Open Field\\\small\textit{(single)}}
& \makecell{CalMS21\\\small\textit{(pair)}} 
& \makecell{Mouse Triplet\\\small\textit{(3 mice)}} 
& \makecell{Open Field\\\small\textit{(single)}}
& \makecell{CalMS21\\\small\textit{(pair)}} 
& \makecell{Mouse Triplet\\\small\textit{(3 mice)}} \\
\midrule
BAMS                  & 0.628\pgstd{0.019} & 0.708\pgstd{0.011} & 0.599\pgstd{0.006} & 0.535\pgstd{0.011} & 0.484\pgstd{0.003} & 0.360\pgstd{0.017} \\
hBehaveMAE            & 0.713\pgstd{0.021} & 0.624\pgstd{0.007} & 0.621\pgstd{0.004} & 0.572\pgstd{0.019} & 0.480\pgstd{0.009} & 0.452\pgstd{0.020} \\
BEAST                 & 0.727\pgstd{0.006} & 0.711\pgstd{0.010} & 0.553\pgstd{0.023} & 0.512\pgstd{0.046} & 0.433\pgstd{0.039} & 0.310\pgstd{0.049} \\
KPMS                  & --                 & --                 & --                 & 0.650\pgstd{0.035} & 0.447\pgstd{0.018} & 0.440\pgstd{0.028} \\
\addlinespace[2pt]
Pose coordinates   & 0.663\pgstd{0.005} & 0.593\pgstd{0.009} & 0.583\pgstd{0.006} & 0.579\pgstd{0.014} & 0.388\pgstd{0.020} & 0.360\pgstd{0.057} \\
Hand-crafted features & 0.709\pgstd{0.007} & \pgbest{0.878\pgstd{0.001}} & 0.684\pgstd{0.012} & 0.589\pgstd{0.034} & \pgbest{0.701\pgstd{0.008}} & 0.469\pgstd{0.030} \\
\addlinespace[2pt]
VideoPrism features   & 0.718\pgstd{0.005} & 0.849\pgstd{0.005} & 0.722\pgstd{0.003} & 0.569\pgstd{0.001} & 0.538\pgstd{0.026} & 0.430\pgstd{0.028} \\
VideoPrism + pose coordinates  & 0.732\pgstd{0.010} & 0.843\pgstd{0.006} & 0.729\pgstd{0.005} & 0.590\pgstd{0.010} & 0.542\pgstd{0.024} & 0.473\pgstd{0.029} \\
\midrule
\textbf{ResiFuse (ours)}
                      & \pgbest{0.796\pgstd{0.009}}
                      & 0.865\pgstd{0.002}
                      & \pgbest{0.784\pgstd{0.002}}
                      & \pgbest{0.655\pgstd{0.011}}
                      & 0.555\pgstd{0.019}
                      & \pgbest{0.537\pgstd{0.024}} \\
\quad\pgbudget{best budget}
                      & \pgbudget{r=75\%}
                      & \pgbudget{r=100\%}
                      & \pgbudget{r=75\%}
                      & \pgbudget{r=25\%}
                      & \pgbudget{r=75\%}
                      & \pgbudget{r=75\%} \\
\bottomrule
\end{tabular}}
\end{table*}

\begin{figure}[t]
  \centering
  \includegraphics[width=\linewidth]{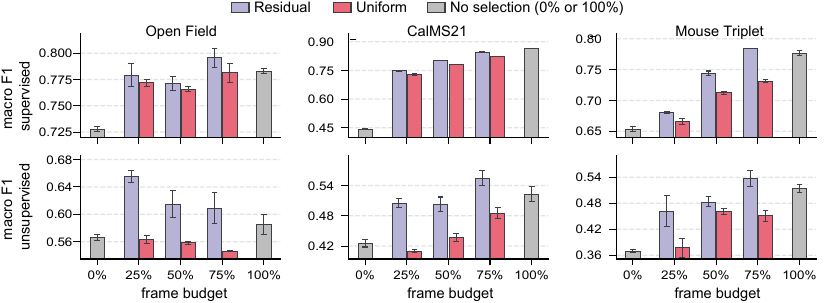}
  \vspace{-10pt}
\caption{\textbf{Residual-guided frame selection consistently outperforms
uniform selection across budgets, datasets, and supervision settings.}
Each panel reports macro F1 on a single dataset
under either supervised fine-tuning (top row) or unsupervised clustering
(bottom row). At intermediate vision budgets (25\%, 50\%, 75\%) we compare
residual selection (purple) against uniform selection (red);
at the endpoints (0\% and 100\%, gray) no selection mode applies. Residual selection
exceeds uniform at every dataset $\times$ setting $\times$ budget,
with the largest gains on Mouse Triplet. A partial budget exceeds full vision (100\%)
except on supervised Open Field (comparable) and CalMS21 (lower).}
 \vspace{-10pt}
  \label{fig:residual-budgets}
\end{figure}

\paragraph{ResiFuse leads on most settings.}
Table~\ref{tab:main} reports macro F1 across the three datasets, under both supervised classification and unsupervised segmentation. ResiFuse achieves the best performance on four of six tasks, with the largest combined gain over the strongest baseline on Mouse Triplet, the most socially complex dataset. This suggests that residual-guided selective fusion is most valuable in complex social settings. ResiFuse already surpasses the strongest baseline with vision on only 25\% (Open Field) and 50\% (Mouse Triplet) of frames (Figure~\ref{fig:residual-budgets}). The exceptions are CalMS21 in both settings, where MARS features hand-engineered for two-mouse recordings lead. Qualitative visualization on Mouse Triplet (Appendix~\ref{app:unsup_viz}) further confirms that our unsupervised segmentations align more closely with ground-truth behaviors than other baselines.

\paragraph{Vision is strong, but naive pose-vision concatenation is not enough.}
VideoPrism features alone outperform every pose-only baseline, except MARS features on CalMS21, yet concatenating them with pose coordinates at every frame changes performance only marginally. Unlocking the complementary value of pose therefore requires more than naive concatenation.

\subsection{Residual-Guided Selection: When and Where It Works}
\label{sec:residual-analysis}

\subsubsection{Across Budgets and Datasets}
\label{sec:residual-budgets}

\paragraph{Residual selection outperforms uniform selection at every dataset and budget, and often matches or exceeds full vision.}
Figure~\ref{fig:residual-budgets} compares residual-guided selection against uniform selection across the three datasets, both supervised and unsupervised settings, at three intermediate budgets (25\%, 50\%, 75\%). In all 18 dataset $\times$ setting $\times$ budget configurations, residual selection wins. Beyond beating uniform selection, a partial budget (<100\%) gives the best result in all three unsupervised settings. In the supervised setting (paired significance tests in Appendix~\ref{app:budget-stats}), the 75\% budget significantly exceeds full vision on Mouse Triplet, is comparable on Open Field, and falls below it on CalMS21, where visual context appears informative on most frames. Residual-selected vision therefore matches or exceeds dense processing in most settings at a lower vision cost.
 
\paragraph{The gap between residual and uniform grows with dataset complexity.}
Comparing the residual-uniform gap across datasets shows a consistent pattern at the 75\% budget: in both supervised and unsupervised settings, the gap is largest on Mouse Triplet, the most socially complex dataset, smallest on Open Field, and intermediate on CalMS21. Intuitively, when behaviors are simple and visual content is relatively uniform across frames, any sufficient subset of frames yields similar information; the choice of which frames to use matters less. As social interactions become more varied and visually intricate, individual frames carry more discriminating content, and identifying the right ones becomes correspondingly more important. The residual signal scales naturally to this regime.


\subsubsection{Across Vision Encoders}
\label{sec:encoder-analysis}

\paragraph{Residual selection generalizes across vision encoders.}
We test whether the benefits of residual-guided selection transfer beyond a single encoder by repeating the budget sweep on Mouse Triplet supervised with five vision encoders: VideoPrism~\citep{zhao2024videoprism}, VideoMAE(V2)~\citep{wang2023videomaev2}, DINOv3~\citep{simeoni2025dinov3}, CLIP~\citep{radford2021learning}, and Qwen3.5 vision encoder~\citep{qwen3.5}. For each encoder, we re-run the full pipeline, including both pretraining and fine-tuning, so that the codebook in each row is shaped by its own encoder's $\mathcal{L}_{\text{pred}}$. For all vision encoders, we extract patch tokens and apply mean pooling to obtain a single feature vector per frame.

\begin{wrapfigure}{r}{0.5\textwidth}
\vspace{-10pt}
\centering
\includegraphics[width=0.5\textwidth]{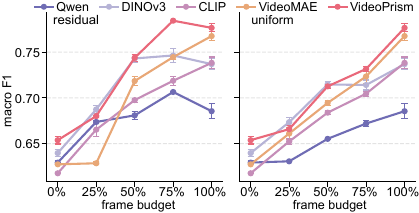}
\caption{\textbf{Residual-guided selection generalizes across vision encoders.} 
Macro F1 on Mouse Triplet supervised across five vision encoders. Residual selection (left) outperforms uniform (right) at most budgets for every encoder.}
\label{fig:encoder-analysis}
\vspace{-10pt}
\end{wrapfigure}

Figure~\ref{fig:encoder-analysis} shows two consistent patterns across the five encoders: residual selection (left) outperforms uniform selection (right) at most budgets, and the best partial-budget result outperforms the 100\% baseline for most encoders. A 50\% budget already recovers 65--97\% of each encoder's maximum gain over no vision. The residual signal is therefore not tied to a particular vision backbone. We then highlight three exceptions to these patterns and explain each in terms of encoder-specific properties. (i) For VideoMAE(V2), uniform selection at the 25\% budget outperforms residual selection at the same budget. (ii) For VideoMAE(V2), the 100\% baseline is also higher than residual at 75\%. Both observations stem from the same design: VideoMAE(V2) outputs one feature per pair of adjacent frames (e.g., 8 features for a 16-frame clip), while our residual selection operates at the per-frame level. Selecting frame $t$ but not frame $t+1$ is not actually possible when the two share a single feature, so residual selection cannot be applied at the granularity it was designed for, and its benefit is reduced. (iii) For CLIP, the 100\% baseline is higher than residual at 75\%. Among the five encoders, CLIP gains the most when features are averaged over neighboring frames (Appendix~\ref{app:encoder-frames}), so adding vision to more frames helps it most. These exceptions reflect properties of the encoders rather than limitations of the residual signal.

\paragraph{Video-pretrained per-frame encoders are best suited for residual-guided selection.}
VideoPrism delivers the strongest residual-driven results: at the 75\% budget it reaches the highest macro F1 across all encoder $\times$ budget combinations, and clearly exceeds its own 100\% baseline. Intuitively, the remaining frames excluded by residual selection are low-residual frames where pose alone is already sufficient; adding their visual features at the 100\% budget brings little new information and can dilute the discriminative signal from the selected high-residual frames. The strong absolute performance reflects VideoPrism's encoder design: it is pretrained on video and produces native per-frame features, providing the per-frame distinctiveness that residual selection requires.

\subsubsection{What Does Residual Select?}
\label{sec:residual-qualitative}
\begin{wrapfigure}{r}{0.5\textwidth}
\centering
\includegraphics[width=0.5\textwidth]{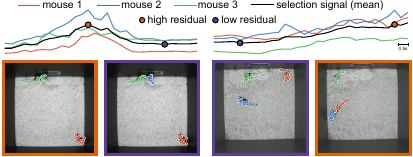}
\caption{\textbf{What residual-guided selection picks.} Two example sequences from Mouse Triplet showing residual signals and corresponding frames at high/low-residual (orange/purple) moments; high-residual frames capture occluded social contact (left) and rapid chase (right), while low-residual frames show tracked, slow-moving postures.}
\label{fig:residual-qualitative}
\end{wrapfigure}
\paragraph{Residual identifies behaviorally meaningful frames without supervision.}
Figure~\ref{fig:residual-qualitative} shows two example sequences from Mouse Triplet, each contrasting a high-residual frame (orange) with a low-residual frame (purple) sampled from the same video. The low-residual frames show three mice clearly tracked with no fast motion, where pose alone provides a clean description of the scene. The high-residual frames capture two distinct failure modes of pose: in the first sequence, green and blue mice are in close contact and blue mouse's keypoints are partially missing due to occlusion, leaving the pose unreliable while the video still contains the interaction (case (ii) in Section~\ref{sec:method-objectives}); in the second sequence, a chase produces overlapping bodies and rapid motion between blue and red mice, where pose alone struggles to characterize the moment (case (i) in Section~\ref{sec:method-objectives}). These qualitative cases are not isolated. A systematic frame-level analysis on the entire Mouse Triplet fine-tuning set (Appendix~\ref{app:quantify}) confirms that residual selection concentrates on three behaviorally meaningful regimes simultaneously: occluded frames, rapid motion, and behavior transitions, all without any supervision.



\subsection{Computational Cost}
\label{sec:cost}
Pretraining uses visual features of all frames and is a one-time cost per dataset that does not depend on the budget. After pretraining, frames are selected from pose alone with one pass of the motion encoder, so in both supervised classification and unsupervised segmentation, visual features are needed only for the selected frames of each recording, during training and at inference; the fusion model itself processes every frame at every budget. We report inference time as the time to process one window on a single GPU, from decoding the selected frames to the output of the fusion model (Table~\ref{tab:cost}). On Mouse Triplet, the 75\% budget reaches the highest macro F1 (0.784 vs.\ 0.777 at full budget) with 76\% of the full-budget inference time, and the 50\% budget retains about two thirds of the best improvement over no vision (0.744, vs.\ 0.654 without vision) with 54\% of the full-budget inference time.

\subsection{Ablation Studies}
\label{sec:ablations}

\begin{wrapfigure}{r}{0.5\textwidth}
\vspace{-10pt}
\centering
\includegraphics[width=0.5\textwidth]{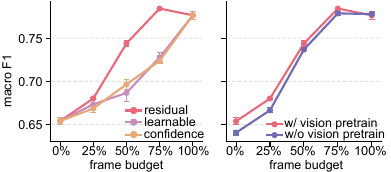}
\caption{\textbf{Ablations on Mouse Triplet supervised.} 
(Left) Selection strategy comparison: residual outperforms keypoint confidence and a learnable MLP selector at every intermediate budget. 
(Right) Without vision in pretraining (motion-only), performance is lower when the fine-tuning budget is low, illustrating that vision cannot be entirely omitted.}
\label{fig:ablations}
\vspace{-10pt}
\end{wrapfigure}

\paragraph{The residual is a stronger selection signal than confidence or learned selectors.}
Figure~\ref{fig:ablations} (left) compares three selection strategies on Mouse Triplet supervised: residual, keypoint confidence, and a learnable MLP selector trained end-to-end with the classification loss. All three share the same motion encoder, vision encoder (VideoPrism), and downstream architecture, and differ only in which frames are selected at intermediate budgets. Residual selection outperforms both alternatives at every intermediate budget, and the advantage widens as budget increases. The gap is intuitive: keypoint confidence reflects tracking quality, not behavioral content, and a frame can be cleanly tracked yet visually uninformative. The learnable selector has only an indirect signal to learn from, since it must extract its decision from a frozen motion latent that was optimized for pose reconstruction rather than for selection, and the classification loss propagates back through hard top-$k$ as a scalar that conflates selection quality with every other component being trained. The residual is grounded in exactly the right signal: it directly measures where the motion encoder's own representation breaks down, which is the property that residual-guided selection ultimately needs.

\paragraph{Cross-modal pretraining shapes residual selection, and selection makes vision affordable when pretraining lacks it.}
Figure~\ref{fig:ablations} (right) compares two motion encoders that differ only in pretraining: one trained with the auxiliary cross-modal objective ($\mathcal{L}_{\text{pred}}$ in Eq.~\ref{eq:pretrain-loss}) and one trained with motion-only objectives. Both use residual selection at supervised fine-tuning. The motion-only pretrain is lower at low fine-tuning budgets (0\%, 25\%), where little vision is supplied, and the two are nearly identical at higher budgets (75\%, 100\%). Two takeaways follow. First, vision must enter the model somewhere: motion-only pretraining can match cross-modal pretraining only when fine-tuning supplies enough vision. Skipping vision at pretraining is recoverable but not free; residual selection is what makes this recovery efficient, by spending the vision budget on the frames that need it most. Second, the F1 gap reflects a concrete mechanism: cross-modal pretraining changes which specific frames the residual selects. A frame-level analysis (Appendix~\ref{app:vision_pretrain_disagreement}) shows the F1 gap and selection differences between two pretrains move together across budgets. We use cross-modal pretraining as our default because it helps under tight fine-tuning budgets, although its gain is small compared with the effect of the budget itself.

\section{Related works}
\label{sec:related}

\textbf{Mouse behavior analysis from pose and video.}
Pose-based methods learn from skeletal trajectories via discrete segmentation~\citep{hsu2021b, weinreb2024keypoint, luxem2022identifying}, multi-scale temporal modeling~\citep{azabou2023relax, stoffl2024elucidating}, or data-efficient embeddings~\citep{sun2021task}, but are limited by the expressiveness of skeletal coordinates alone. Video-based methods instead operate on raw pixels: masked autoencoding~\citep{he2022masked, tong2022videomae} transfers poorly to multi-animal settings~\citep{sun2023mabe22}, motivating domain-specific models~\citep{sun2022self, wang2025self} and adapted video foundation models~\citep{sun2024video, xu2025mousegpt, liu2025castle}. \textbf{Selective vision processing.}
In animal behavior analysis, existing pose-video approaches rely on dense feature concatenation across all frames~\citep{segalin2021mouse, goodwin2024simple, sun2023mabe22, kozlova2025dlc2action}.
Selective vision processing has instead been explored in efficient video recognition, where a policy trained with the recognition objective decides which frames or modalities to process~\citep{gao2020listen}, in video language models through learned keyframe selectors~\citep{tang2025adaptive, wang2025videotree, zhu2025focus}, and in video anomaly detection through reconstruction error as an unsupervised frame-level signal~\citep{gong2019memorizing, park2020learning}. Learned selectors are tied to a specific task or query, and reconstruction error is measured in input space, where it stays low on occluded frames (Appendix~\ref{app:residual-recon}).
Our work differs from these lines: we use the quantization residual from the latent space of a pretrained motion encoder, a parameter-free by-product of training computed from pose before any frame is decoded, as a selection signal for pose-video fusion.

\section{Discussion}
\label{sec:conclusion}
We presented ResiFuse, a framework that integrates pose and video for mouse behavior analysis through residual-guided selective fusion. Across three datasets and both supervised classification and unsupervised segmentation, residual-guided selection outperforms uniform selection at every intermediate budget, and a partial budget matches or exceeds full-vision processing in most settings. Because frames are selected from pose alone, visual features are needed only for the selected frames. Beyond performance, the learned codebook itself captures interpretable social-contact patterns (Appendix~\ref{app:codebook_viz}). ResiFuse has several limitations. It is trained from scratch per dataset rather than as a foundation model; jointly pretraining across datasets and tasks is a natural next step that we expect to scale well given the parameter-free nature of residual selection. Pretraining still uses visual features of all frames, the vision budget is a single ratio fixed per dataset rather than adapted to each recording.

\section*{Broader Impacts}
\label{app:broader_impacts}

This work focuses on automated analysis of laboratory mouse behavior. The primary positive impact is in scientific practice: more accurate behavior quantification can improve reproducibility, accelerate basic neuroscience research, and potentially reduce the number of animals needed for statistical power in preclinical studies, consistent with the Reduction principle of the 3Rs framework. The method is specific to laboratory mouse recordings and does not apply to human subjects, so we do not foresee direct negative societal impacts.

\section*{Acknowledgement}

This work was supported by the National Institute of Mental Health (NIMH) of the National Institutes of Health (NIH) (RF1MH140429) and the National Science Foundation (NSF) (2336055).

\bibliographystyle{plainnat}
\bibliography{cited}

\newpage


\appendix

\section{Implementation Details}
\label{app:implementation}
 
\subsection{Architecture}
 
\paragraph{Tokenizer.}
Three stride-1 residual convolutional blocks (no temporal downsampling; 256 channels for Mouse Triplet and CalMS21, 128 for Open Field) are followed by a projection to the RVQ latent dimension $D = 128$ and by temporal self-attention over the window (4 heads; two layers for Mouse Triplet and CalMS21, one for Open Field). The convolutions give each latent a receptive field of 21 frames and the self-attention adds context from the whole window, so the residual reflects temporal structure rather than a single static pose.
The decoder is a mirrored structure that reconstructs the motion features.
 
\paragraph{RVQ.}
Two-level residual vector quantization. Codebook sizes are set a priori from the number of animals rather than tuned on accuracy: 512 and 512 entries for Mouse Triplet (three mice), 512 and 128 for CalMS21 (two mice), and 256 and 64 for Open Field (one mouse).
Codebooks are updated via Exponential Moving Average (EMA) with decay $0.99$ and a commitment weight $\beta = 0.25$.
Dead-code revival is enabled to keep the codebook well-utilized. After pretraining, all entries remain in use on every dataset, with effective codebook sizes (the exponential of the usage entropy) of 250 and 373 on Mouse Triplet, 71 and 119 on CalMS21, and 25 and 31 on Open Field. Because frame selection uses only the ranking of residuals within a budget rather than their magnitude, it does not depend on the overall scale of quantization error set by the codebook size.
 
\paragraph{Context and social Transformers.}
Both are 2-layer self-attention Transformers with 8 heads, model dimension $D$, and feed-forward dimension $4D$.
The context Transformer operates over the temporal dimension; the social Transformer operates over the animal dimension at each timestep.
 
\paragraph{Vision projector.}
The vision projector is a single linear layer followed by LayerNorm.

\paragraph{Classifier.}
Since behavior classification at each frame depends on temporal context, we use a sequence classifier rather than a static per-frame one. It consists of a single-layer BiGRU with hidden size 256 and a residual connection, followed by a 2-layer MLP with hidden size 256 that maps to the number of behavior classes. The same classifier is applied to ResiFuse and all supervised baselines. Because each vision encoder outputs features frame by frame (one feature per pair of frames for VideoMAE), this BiGRU provides all temporal context for the visual features of the image encoders (CLIP, DINOv3, Qwen3.5); the video encoders additionally mix information across frames within each input clip, but not across clips.
 
\subsection{Pretraining details.}
\label{app:pretrain-details}
 
\paragraph{Input features.}
For each animal, per-frame motion features consist of joint coordinates $[x, y]$ and per-joint confidence scores from the pose estimator. Coordinates are divided by the frame width and height for Mouse Triplet and CalMS21; for Open Field, the dataset provides egocentric keypoints (Appendix~\ref{app:experiment}). On Mouse Triplet and CalMS21, we further append features of the $M$ nearest other animals ($M=2$ and $1$, respectively): for each neighbor, the distance between body centers divided by the frame size, the sine and cosine of its direction, and its mean keypoint confidence, adding $4M$ dimensions.
 
\paragraph{Pretraining objectives.}
We train with four losses:
\begin{equation}
\mathcal{L}_{\text{pretrain}} = \mathcal{L}_{\text{mask}} + \lambda_{\text{recon}} \mathcal{L}_{\text{recon}} + \lambda_{\text{vq}} \mathcal{L}_{\text{vq}} + \lambda_{\text{pred}} \mathcal{L}_{\text{pred}},
\end{equation}
where:
 
$\mathcal{L}_{\text{mask}}$: loss of reconstruction of the masked pose. 
 
$\mathcal{L}_{\text{recon}}$: RVQ reconstruction loss. MSE between the original motion features and the decoded $z_q$.
 
$\mathcal{L}_{\text{vq}}$: the standard loss that stabilizes codebook learning.
 
$\mathcal{L}_{\text{pred}}$: multi-horizon visual prediction loss. We predict the frozen vision encoder's features at three future horizons $\mathcal{H} = \{4, 8, 16\}$ (in downsampled frames) using three independent prediction heads, and sum per-horizon cosine similarity losses with weights $[1.0, 0.7, 0.4]$.
 
Loss weights: $\lambda_{\text{recon}} = 0.1$, $\lambda_{\text{vq}} = 0.1$, and $\lambda_{\text{pred}} = 0.05$ / $0.2$ / $0.1$ for Mouse Triplet / CalMS21 / Open Field. The weight $\lambda_{\text{pred}}$ is kept small so that this loss acts as an auxiliary regularizer on the codebook rather than dominating the optimization.

\paragraph{Social asymmetric M-MAE masking.}
Instead of applying a uniform random mask across all animals, we use an asymmetric masking strategy designed to make the masked reconstruction objective align with the multi-animal structure of the problem.
At each training step, one animal is randomly selected as the \emph{target}; the remaining animals serve as \emph{context}.
The target's frames are masked at a high rate ($\rho_{\text{target}} = 0.5$), while context animals are masked lightly ($\rho_{\text{context}} = 0.15$).
Reconstructing the heavily-masked target then requires the network to look across animals through the social attention layers, rather than relying solely on the target's own sparse context.
 
For each animal, the masking budget is split equally between two patterns:
frame-level i.i.d.\ dropout (Bernoulli sampling at the per-frame masking rate), and block dropout (random contiguous blocks of length uniformly sampled from $[4, 8]$ until the block-level budget is reached).
The two patterns complement each other: pure frame-level masking is often solvable by local interpolation, while pure block masking is too hard; mixing them balances fine-grained dynamics and longer-range structure.
Frames that are invalid in the input (padding or failed detections) are always excluded from masking and from the reconstruction loss.
 
\paragraph{Pretraining hyperparameters.}
AdamW optimizer with initial learning rate $1 \times 10^{-4}$, weight decay $0.01$, batch size 64.
Cosine annealing schedule with $T_{\max} = 150$ epochs and $\eta_{\min} = 1 \times 10^{-5}$.
Training is done with Distributed Data Parallel on 4 NVIDIA L40 GPUs.
 
\subsection{Fine-tuning details.}
 
\paragraph{Supervised fine-tuning.}
\label{app:sup-details}
All pretrained components of the motion encoder (tokenizer, RVQ, context Transformer, social Transformer) are frozen.
Trainable parameters: vision projector, motion/fused feature normalization layers, a classifier. In our experiments, visual features are precomputed once for all frames and masked according to $\mathcal{S}$; fine-tuning time is therefore independent of the budget, and the budget reduces only the cost of decoding and encoding frames.
 

We use AdamW with learning rate $1 \times 10^{-4}$, weight decay $0.01$, 50 epochs.
Loss is binary cross-entropy with positive-class reweighting for multi-label tasks, and cross-entropy for single-label tasks.
 
\paragraph{Unsupervised fine-tuning (DEC).} \label{app:unsup-details}
Same frozen backbone as in supervised fine-tuning. DEC operates directly on the fused representation $h_t \in \mathbb{R}^D$ from Section.~\ref{sec:ht} ($D=128$). Trainable parameters: vision projector, feature normalization layers, and $K$ cluster centers $\{\mu_k\}_{k=1}^K \in \mathbb{R}^D$. The visual term of $h_t$ is centered by subtracting its mean over selected training frames, computed once before initialization and then kept fixed. Cluster centers are initialized by running batched K-means on 50{,}000 $h_t$ vectors sampled uniformly from the training set for 50 iterations, then optimized jointly with the other trainable parameters. To ensure consistency with existing baselines, we initialize $K=50$ cluster centers for Mouse Triplet and CalMS21, and $K=30$ for Open Field.

The DEC objective uses a Student-$t$ soft assignment and a sharpened target distribution~\citep{xie2016unsupervised}:
\begin{equation}
q_{tk} = \frac{(1 + \|h_t - \mu_k\|^2)^{-1}}{\sum_{k'} (1 + \|h_t - \mu_{k'}\|^2)^{-1}}, \quad
p_{tk} = \frac{q_{tk}^2 / \sum_{t'} q_{t'k}}{\sum_{k'} q_{tk'}^2 / \sum_{t'} q_{t'k'}},
\end{equation}
minimized through $\mathcal{L}_{\text{DEC}} = \text{KL}(p \,\|\, q)$ with $p$ detached.

To prevent representation collapse, we add a VICReg-style regularization~\citep{bardes2021vicreg} on time-adjacent frames (shift $= 1$). Given two views $f_a = h_t$ and $f_b = h_{t+1}$, the regularization has three terms: invariance $\mathcal{L}_{\text{inv}} = \text{MSE}(f_a, f_b)$, variance $\mathcal{L}_{\text{var}} = \frac{1}{D} \sum_d \text{ReLU}(\gamma - \sigma_d(f_a))$ with $\gamma = 1$, and covariance $\mathcal{L}_{\text{cov}} = \frac{1}{D} \sum_{i \neq j} \text{Cov}(f_a)_{ij}^2$, where $\sigma_d$ denotes the standard deviation of dimension $d$ over the batch and $\text{Cov}(f_a)$ is the covariance matrix of $f_a$ across the batch. The total unsupervised objective is
\begin{equation}
\mathcal{L}_{\text{unsup}} = \mathcal{L}_{\text{DEC}} + w_{\text{inv}} \mathcal{L}_{\text{inv}} + w_{\text{var}} \mathcal{L}_{\text{var}} + w_{\text{cov}} \mathcal{L}_{\text{cov}}.
\end{equation}
We set $w_{\text{inv}}=0.5$, $w_{\text{var}}=1.0$, and $w_{\text{cov}}=0.1$ on all datasets and for all methods.
 
Two learning rates: backbone (vision projector, normalization layers) at $1 \times 10^{-4}$, cluster centers at $1 \times 10^{-4}$.
AdamW, 30 epochs, no scheduler. At the start of each epoch, the target distribution $p$ is computed on the full training set and kept fixed within the epoch, and empty clusters are reseeded at a random member of the largest cluster. At evaluation time, each cluster is mapped to its majority class on the labeled training split (count-based mapping by default; F1-based mapping as an alternative for imbalanced classes).

\newpage
\section{Experiment Details}
\label{app:experiment}

\subsection{Data preprocessing}
 
\paragraph{Temporal subsampling.}
The input sequence length $L$ used throughout the method is obtained by subsampling the raw video frames by a factor of $4$ at the data loading stage.
The motion encoder itself does not perform any further temporal downsampling.

\paragraph{Per-animal video cropping for vision features.}
Vision features are extracted on per-animal video crops rather than on full frames.
For each frame, we estimate the center of each animal from its keypoints and crop a window around that center.
The crop size is chosen to be substantially larger than the animal itself, so that the crop preserves the surrounding context: this ensures that important visual cues such as social interactions between animals or the proximity of environmental objects (e.g., walls, feeders) are visible within the crop, rather than being truncated at the boundary.
The frozen vision encoder is applied to each crop independently, producing a per-animal visual feature for each frame.

\paragraph{Open field Preprocessing} For the open field dataset \citep{sturman2020deep}, the data are egocentric: both the videos and the estimated keypoints are provided in an egocentric coordinate frame. We construct pretraining datasets using a sliding window of 256 frames with a stride of 128. During fine-tuning, we use 256-frame windows with no overlap to ensure each frame is evaluated exactly once, and each window is required to contain labels. Windows shorter than the specified length are zero-padded, and padded frames are masked out during training and loss computation. This applies to all methods for fair comparison. No frames are discarded owing to the high quality of the pose in this dataset. Both pretraining and fine-tuning use the same 50\%/10\%/40\% train/val/test split. The dataset annotates supported and unsupported rearing separately; we merge them into one rearing class because they differ mainly in proximity to the arena wall, which the egocentric keypoints given to all pose-based methods cannot represent but the video shows, so keeping them separate would favor methods with video access. With the two rearing classes kept separate, ResiFuse reaches a three-class macro F1 of $0.837\pm0.010$.

\paragraph{CalMS21 Preprocessing} For the CalMS21 dataset \citep{sun2021multi}, we use the official Task 1 split, which contains 89 videos with MARS-tracked keypoints (7 per mouse) and per-frame behavior labels (attack, investigation, mount, other). The keypoints contain no missing detections, so no additional pose cleanup is required; one anomalously short clip (3.2~s) is excluded, leaving 88 videos. We create pretraining datasets using a sliding window of 256 frames with a stride of 128, and during fine-tuning, the input sequence is split into non-overlapping windows of 256 frames. Windows shorter than the specified length are zero-padded, and padded frames are masked out during training and loss computation. This applies to all methods for fair comparison. Both pretraining and fine-tuning follow the predefined Task 1 split.

\paragraph{Mouse Triplet Preprocessing} For the mouse triplet dataset \cite{sun2023mabe22}, we use the original pose and videos from the \emph{mouse submission folder}, which contains approximately 1,800 videos. We manually annotate a subset of videos with high-quality pose to train a DeepLabCut model~\citep{mathis2018deeplabcut}, which is then used to re-estimate pose for all videos. All methods (baselines and ResiFuse) are evaluated on these re-estimated pose, ensuring fair comparison; baseline numbers therefore differ from those originally reported on the raw Mouse Triplet pose. Frames with low confidence scores of all mice are discarded. We create the pretraining dataset using a sliding window of 300 frames with a stride of 150, and the fine-tuning dataset using a sliding window of 200 frames with no overlap, requiring label inclusion. Windows shorter than the specified length are zero-padded, and padded frames are masked out during training and loss computation. This applies to all methods for fair comparison. Pretraining uses 70\%/20\%/10\%, fine-tuning uses 50\%/10\%/40\%; no video in the fine-tuning test set appears in the pretraining set.

\begin{figure}[!thp]
  \centering
  \includegraphics[width=0.75\linewidth]{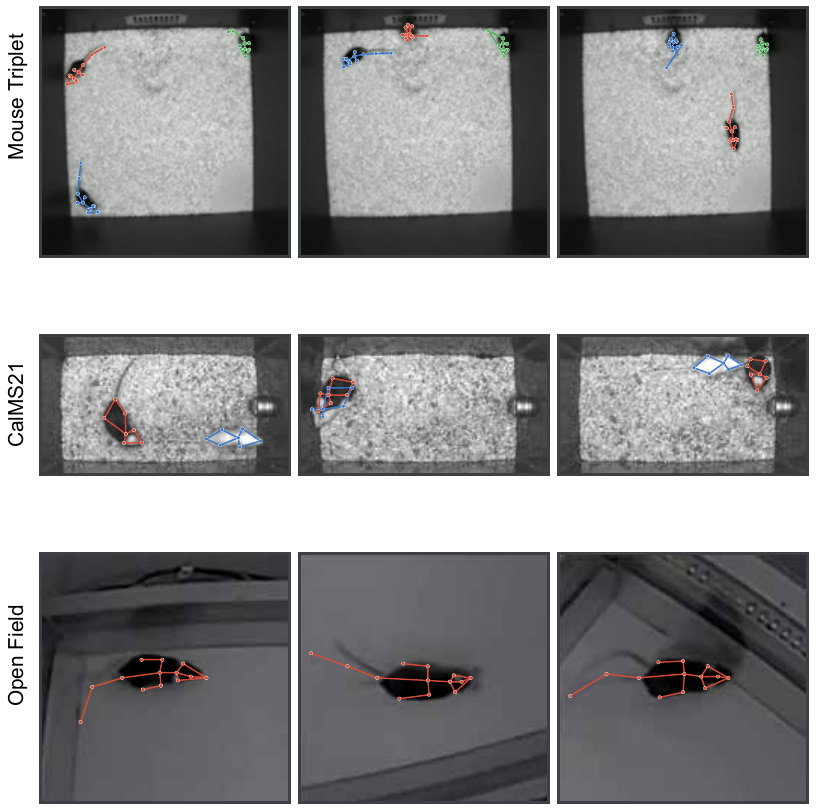}
\caption{Example frames for Mouse Triplet, CalMS21, and Open field mouse datasets.}
\label{fig:example}
\end{figure}

See example frames for three mouse datasets in Figure~\ref{fig:example}.

\subsection{Baseline settings}

\paragraph{BAMS settings.}
We use the official BAMS implementation~\citep{azabou2023relax} as our self-supervised baseline, utilizing its default dual-branch TCN encoder to produce 128-d embeddings. Pretraining (200~epochs, AdamW) follows the HoA+BYOL objective with dataset-tuned targets: we use rotation-invariant speed magnitudes $\|v_t\|=\sqrt{(\Delta x_t)^2+(\Delta y_t)^2}$ for Mouse Triplet and CalMS21 to capture social dynamics, while retaining per-axis displacements for Open Field. The HoA window and bin count are also dataset-tuned: 32/16~frames (pretrain/finetune) with 16~bins on Mouse Triplet and CalMS21 to span second-scale social dynamics, and 8/8~frames with 24~bins on Open Field so the histograms stay dense over the sub-second timescale of grooming. For supervised fine-tuning, we freeze the encoder and train a BiGRU-MLP head for 100~epochs, matching the architectures and hidden sizes used by ResiFuse for direct comparability. Loss functions are multi-label BCE with $\texttt{pos\_weight}$ for Mouse Triplet and CalMS21 and weighted multi-class softmax for Open Field. For unsupervised fine-tuning, we project pretrained features onto a 128-d space and perform clustering. To ensure direct comparability with ResiFuse, we employ the same Deep Embedded Clustering loss \citep{xie2016unsupervised} and variance regularization. We set $50$ clusters for Mouse Triplet, $50$ for CalMS21, and $30$ for Open Field.

\paragraph{BEAST settings.}
We use the official BEAST implementation~\citep{wang2025self} where a ViT-B/16 encoder initialized from \texttt{facebook/vit-mae-base} and pretrained with masked autoencoding on unlabeled video frames using a 75\% masking ratio, top-down rotation/flip/crop augmentation, AdamW, cosine scheduling, and a learning rate of $1.5\times10^{-4}$. For Mouse Triplet, we pretrain for 400 epochs and use the best checkpoint to extract frozen 768-d visual features; for Open Field, we use the same protocol but pretrain for 200 epochs. For CalMS21, we use the released BEAST ViT-B/16 checkpoint from \texttt{paninski-lab/vitb16-beast-calms21}. For supervised classification, we freeze the encoder and train BEAST's native segmentation head on its patch embeddings: multi-head attention pooling (8 heads) over patch tokens, concatenated with frame-to-frame differences and followed by a two-block dilated TCN (hidden size 64, kernel 17), trained for 100 epochs with AdamW (learning rate $10^{-3}$, weight decay $10^{-2}$). For Mouse Triplet, patch tokens are projected to 128 dimensions before pooling and the kernel is 9 to match the shorter windows. BEAST does not report its final TCN hyperparameters, so we searched its candidate ranges (kernel $\{9, 17, 33\}$, hidden size $\{16, 32, 64, 96\}$) and found the configuration above to be close to the best. Two protocol differences remain relative to the original BEAST evaluation, which reports $0.81\pm0.01$ for a single model on CalMS21: on CalMS21, all methods in our benchmark use 64-frame windows at 4$\times$ temporal subsampling (256 raw frames) instead of 500 frames at the full frame rate, and we report single models rather than the five-seed logit ensemble behind BEAST's best result. For reference, mean-pooling the patch embeddings and training the BiGRU-MLP head used by the other baselines gives 0.669, 0.473, and 0.514 on Open Field, CalMS21, and Mouse Triplet. For unsupervised fine-tuning, we project mean-pooled patch features onto a 128-d space and perform clustering. To ensure direct comparability with ResiFuse, we employ the same Deep Embedded Clustering loss \citep{xie2016unsupervised} and variance regularization. We set $50$ clusters for Mouse Triplet, $50$ for CalMS21, and $30$ for Open Field.

\paragraph{hBehaveMAE settings.}
We use the official hBehaveMAE implementation~\citep{stoffl2024elucidating}. We train hBehaveMAE with a three-stage hierarchical Hiera encoder, using stage depths $(3,4,5)$ and embedding dimensions $(128,192,256)$. For all three datasets, we pretrain for 200 epochs with masked reconstruction, random masking ratio 0.75, normalized reconstruction loss, AdamW-style base learning rate $1.6\times10^{-4}$, 20 warmup epochs, and a lightweight multi-scale decoder with 128-d embeddings and one transformer block. For supervised fine-tuning, we freeze the pretrained encoder and train only a BiGRU-MLP frame-level head for 100 epochs using AdamW with learning rate $10^{-4}$, weight decay $10^{-2}$, matching the architectures and hidden sizes used by ResiFuse for direct comparability. We sweep hierarchical feature levels $L\in\{0,1,2\}$, selecting $L=0$ for Mouse Triplet and $L=1$ for CalMS21 and Open Field based on the performance. Mouse Triplet and CalMS21 are optimized with multi-label BCE using dataset-computed \texttt{pos\_weight}, whereas Open Field uses weighted multi-class cross-entropy with inverse-frequency class weights. For unsupervised evaluation, we extract frozen per-frame features from the selected hierarchical level for each dataset. Then, we project pretrained features onto a 128-d space and perform clustering. To ensure direct comparability with ResiFuse, we employ the same Deep Embedded Clustering loss \citep{xie2016unsupervised} and variance regularization. We set $50$ clusters for Mouse Triplet, $50$ for CalMS21, and $30$ for Open Field.

\paragraph{Keypoint-MoSeq settings.}
We use the official Keypoint-MoSeq implementation~\citep{weinreb2024keypoint}. We evaluate Keypoint-MoSeq as a purely unsupervised behavior-discovery baseline on Mouse Triplet, CalMS21, and Open Field. Unlike representation-learning baselines, Keypoint-MoSeq is fit directly to raw animal keypoints and infers discrete behavioral syllables with an autoregressive hidden Markov model; no supervised fine-tuning, DEC objective, or variance regularization is used. For all datasets, we use keypoint confidences to mask invalid observations, and fit the model for $200$ iterations followed by $200$ inference iterations. The dataset-specific Keypoint-MoSeq state budgets are $50$ syllables for Mouse Triplet, $50$ for CalMS21, and $20$ for Open Field, with a shared latent pose dimension of $10$ and AR order $3$. For evaluation, inferred syllables are converted to semantic behavior classes by a train-split majority mapping. 

\begin{table}[!thp]
\centering
\caption{Supervised macro F1 of pose featurizations on CalMS21. Pose coordinates, MARS, and ResiFuse also appear in Table~\ref{tab:main}.}
\label{tab:calms21-pose}
\small
\begin{tabular}{l c}
\toprule
Input & Macro F1 \\
\midrule
Pose coordinates & 0.593\pgstd{0.009} \\
Pose + velocity~\citep{blau2024study} & 0.744\pgstd{0.005} \\
TREBA~\citep{sun2021task} & 0.783\pgstd{0.003} \\
MARS~\citep{segalin2021mouse} & 0.878\pgstd{0.001} \\
\midrule
ResiFuse (ours) & 0.865\pgstd{0.002} \\
\bottomrule
\end{tabular}
\end{table}

\paragraph{Pose-feature and VideoPrism-feature baselines.}
These supervised baselines feed per-frame features to the same BiGRU-MLP classifier as ResiFuse, with the same loss, windows, and splits. For unsupervised segmentation, the same features are projected to 128 dimensions and clustered with the same DEC objective, regularization, and number of clusters as the other baselines. \emph{Pose coordinates} use the normalized keypoint coordinates (Appendix~\ref{app:pretrain-details}); \emph{VideoPrism features} and \emph{VideoPrism + pose coordinates} use the per-frame VideoPrism features alone or concatenated with these coordinates. \emph{Hand-crafted features} are dataset-specific: for Open Field, the features provided with the dataset~\citep{sturman2020deep} and their velocities (42 dimensions); for CalMS21, the official 158-dimensional per-mouse MARS features~\citep{segalin2021mouse}; and for Mouse Triplet, 214 group-level geometric features computed with the hand-feature code of a MABe22 baseline~\citep{sun2023mabe22}, using arena corners detected in each video. On CalMS21, we additionally evaluate two featurizations that are available only for this dataset (Table~\ref{tab:calms21-pose}): coordinates concatenated with their frame-to-frame velocities (\emph{Pose + velocity})~\citep{blau2024study}, and the official 32-dimensional TREBA trajectory embeddings~\citep{sun2021task}. All engineered and learned pose features are z-scored with training-set statistics.

\clearpage
\section{Additional Results}

\subsection{Unsupervised Segmentation for Mouse Triplet}
\label{app:unsup_viz}

\begin{figure}[!thp]
  \centering
  \includegraphics[width=\linewidth]{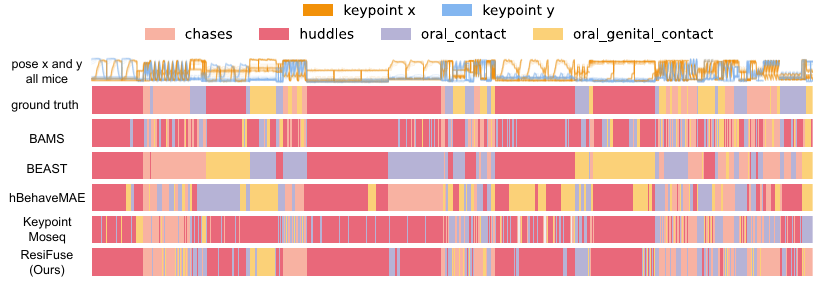}
\caption{\textbf{Qualitative comparison of unsupervised segmentation on Mouse Triplet.} 
Test sequences are concatenated with background frames removed. From top to bottom: pose trajectories of all three mice, ground-truth behavior labels, and predicted segmentations from each method. ResiFuse produces temporally stable segments aligned with ground-truth boundaries, while BAMS and Keypoint-MoSeq exhibit frequent rapid switching within behavioral bouts.}
\label{fig:unsup-viz}
\end{figure}

Figure~\ref{fig:unsup-viz} visualizes the unsupervised segmentation results on the Mouse Triplet fine-tuning test set. We concatenate all test sequences after removing background frames, then plot the ground-truth behavior labels (\emph{chase}, \emph{huddle}, \emph{oral contact}, \emph{oral-genital contact}) and the cluster assignments produced by each unsupervised method, after each method's clusters are mapped to behavior classes via majority-vote on the labeled training split. The top row plots the raw $x$ and $y$ keypoint trajectories for context.

Two patterns are visible. First, BAMS and Keypoint-MoSeq show frequent rapid switching within stretches that the ground truth assigns to a single behavior, an over-segmentation pattern in which the cluster assignment changes many times within what is behaviorally a single bout. BEAST, hBehaveMAE, and ResiFuse (ours) produce smoother segmentations. Second, ResiFuse produces segmentations that are temporally stable and align closely with the ground-truth boundaries: its predicted bouts coincide with ground-truth bouts in both location and length, with switches concentrated near actual transitions rather than within stable behaviors. This visual difference is consistent with the macro F1 numbers reported in Table~\ref{tab:main}, where ResiFuse outperforms all baselines on the unsupervised setting of Mouse Triplet. Appendix~\ref{app:cluster-composition} further shows that these clusters group frames by behavior rather than by whether they received visual features.

\subsection{Clusters Group Frames by Behavior, Not by Visual Input}
\label{app:cluster-composition}

\begin{table}[!thp]
\centering
\caption{Composition of unsupervised clusters on the Mouse Triplet test set at a 75\% budget, pooled over 3 runs. Non-empty clusters are grouped by their dominant behavior. Frames with vision is the share of frames in these clusters that received visual features; purity is the share of frames carrying the dominant behavior.}
\label{tab:cluster-composition}
\small
\begin{tabular}{l c c c}
\toprule
Dominant behavior & Clusters & Frames with vision (\%) & Purity \\
\midrule
huddles & 30 & 47 & 0.81 \\
chases & 8 & 99 & 0.59 \\
oral contact & 9 & 87 & 0.49 \\
oral-genital contact & 1 & 24 & 0.43 \\
\bottomrule
\end{tabular}
\end{table}

Because only selected frames carry visual features, the fused representation could in principle separate frames with and without vision instead of grouping them by behavior. We check this on Mouse Triplet by assigning each non-empty cluster to its dominant behavior (Table~\ref{tab:cluster-composition}). The clusters follow behavior rather than visual input: huddle clusters draw 47\% of their frames from frames with vision and the rest from pose-only frames, and across behaviors this share ranges from 24\% to 99\% without the clusters splitting along it. Chase clusters are almost entirely visual because chase frames have the highest residuals, so nearly all of them are selected at this budget.

\subsection{Codebook Visualization}
\label{app:codebook_viz}

The RVQ in motion encoder uses two codebooks: the level-1 codebook (cb1) encodes the coarse-grained motion state of each frame, while the level-2 codebook (cb2) quantizes the residual of cb1 to refine the representation. We focus this visualization on cb1 because it captures the behavioral primitives, while cb2 refines the representation within each cb1 cluster without defining new behavioral primitives.

To verify that the encoder learns interpretable temporal structure, we visualize \emph{$4$-grams}, ordered tuples of four consecutive cb1 tokens, treated as 4-frame motion motifs, akin to the reusable chunks in cognitive models of gesture learning~\citep{he2024cognitive}. After encoding the entire pretrain set with the frozen motion encoder, we enumerate all token $4$-grams across $(\text{window}, \text{animal})$ pairs, exclude self-loop $4$-grams (all four tokens identical, which encode token persistence rather than motion), and keep only $4$-grams with at least $50$ occurrences for statistical reliability. Each remaining $4$-gram is scored by its mean per-frame keypoint displacement averaged over all its instances, and the top $5$ motion-ranked $4$-grams are selected; we use motion-ranked instead of frequency-ranked because the most frequent $4$-grams correspond to stationary patterns (mice barely moving) rather than informative transitions. For each selected $4$-gram, we pick the single occurrence with the highest average keypoint visibility and display a $6$-frame strip ($1$ context frame, $4$ core frames matching the codebook token sequence, $1$ context frame; core frames bordered in red), with the focused mouse highlighted in red and the other two mice rendered as scene context.

\begin{figure}[!thp]
  \centering
  \includegraphics[width=\linewidth]{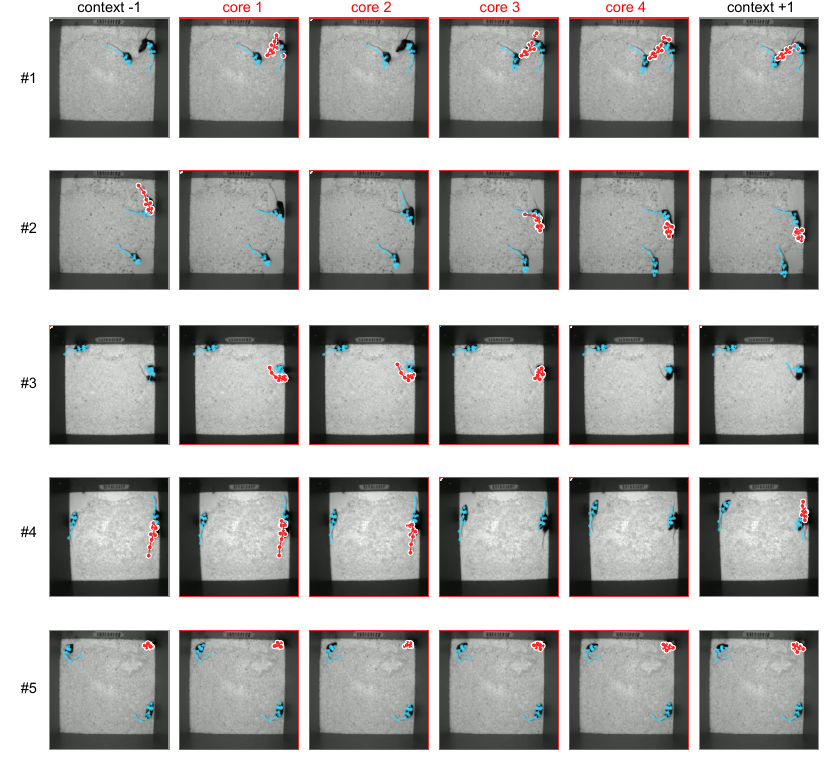}
\caption{\textbf{Visualization of top motion-ranked codebook $4$-grams.}
Each row shows one $4$-gram (an ordered tuple of four consecutive codebook tokens) as a $6$-frame strip: $1$ context frame, $4$ core frames matching the token sequence (red border), and $1$ context frame. The focused mouse is rendered in red; the other two mice provide scene context.}
\label{fig:codebook-viz}
\end{figure}

Figure~\ref{fig:codebook-viz} shows the result. Each row corresponds to one $4$-gram and depicts a multi-frame motion pattern: rows $1$--$4$ capture social-contact motifs such as oral-oral contact and oral-genital contact between two mice, while row $5$ corresponds to a position-driven pattern in which the focused mouse remains stationary at a specific location (e.g., in a corner) rather than performing a distinct action. Two further observations are worth noting. First, several top-ranked $4$-grams (rows $2$--$4$) include core frames in which one mouse's keypoints are partially missing due to occlusion, yet the codebook tokens consistently group these clips together as a coherent action sequence. We hypothesize that the cross-modal pretraining objective 
contributes to this robustness, since aligning the motion latent with visual features 
during pretraining encourages the codebook to capture behavioral structure that is 
not entirely tied to complete pose tracking. Second, the existence of position-driven patterns (such as row $5$) suggests that the codebook is not perfectly disentangled from spatial layout, and not every codebook entry corresponds to a behaviorally meaningful motif. Overall, these visualizations support our use of the codebook structure (and its residual) as a behaviorally informative signal, while acknowledging that a portion of the codebook capacity is consumed by location-specific patterns.

\subsection{Quantitative analysis of residual-selected frames}
\label{app:quantify}

\begin{figure}[!thp]
  \centering
  \includegraphics[width=\linewidth]{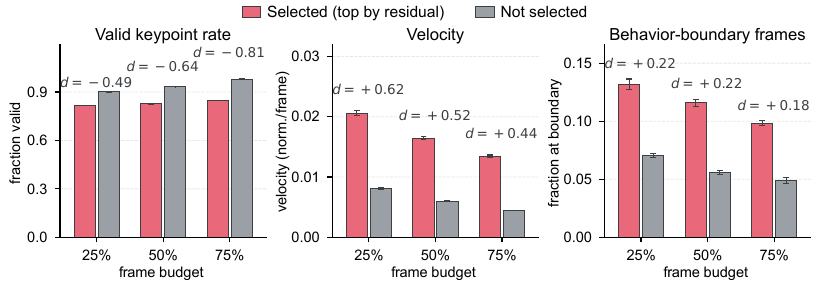}
\caption{\textbf{Residual-selected frames systematically differ from the rest in three behaviorally meaningful ways.} On Mouse Triplet, at each budget ($25\%$, $50\%$, $75\%$), residual-selected frames (pink) are compared with the rest (gray) on three metrics. Bar height is the group mean; whiskers are SEM; Cohen's $d$ is annotated above each pair. \textbf{(Left) Valid keypoint rate}: selected frames have systematically lower keypoint validity ($d$ from $-0.49$ to $-0.81$), indicating residual selection concentrates on frames where pose tracking is unreliable. \textbf{(Middle) Inter-frame velocity}: selected frames are $2.5$ to $3.1\times$ faster ($d$ from $+0.44$ to $+0.62$), indicating residual selection concentrates on rapid-motion frames. \textbf{(Right) Behavior-boundary rate}: selected frames roughly twice as often fall at transitions between consecutive behavior labels ($d$ from $+0.18$ to $+0.22$), indicating residual selection also concentrates on the moments where the model must decide between competing labels.}
\label{fig:residual-quantitative}
\end{figure}

To complement the qualitative examples in Section~\ref{sec:residual-qualitative}, we quantify what high-residual frames look like across the entire Mouse Triplet fine-tune set. For each frame we compute three metrics that are sensitive to behaviorally meaningful events: \emph{(left) valid keypoint rate}, the fraction of tracked keypoints with non-zero confidence (lower indicates more occlusion or tracking failure); \emph{(middle) inter-frame velocity}, the mean keypoint motion magnitude between consecutive frames in normalized coordinates; and \emph{(right) behavior-boundary rate}, the fraction of frames located at a transition between two consecutive behavior labels. At each of three budgets ($25\%$, $50\%$, $75\%$), we split frames into \emph{selected} (top-budget by residual) and \emph{not selected}, and compare the means of each metric across the two groups (Figure~\ref{fig:residual-quantitative}).

The three metrics support the same conclusion (effect sizes reported as Cohen's $d$, the gap between groups in standard-deviation units; $|d|\approx 0.2$ is conventionally a ``small'' effect, $0.5$ ``medium'', and $0.8$ ``large''). Selected frames have systematically lower keypoint validity ($d$ from $-0.49$ to $-0.81$), $2.5$ to $3.1\times$ higher inter-frame velocity ($d$ from $+0.44$ to $+0.62$), and roughly twice as often coincide with behavior transitions ($d$ from $+0.18$ to $+0.22$). Residual selection therefore preferentially identifies frames in which (i) animals are moving rapidly or one behavior is transitioning into another, both of which can leave the pose insufficient to characterize the moment, and (ii) pose tracking is unreliable, where the visual content cannot be inferred from pose alone. These three regimes match the first two cases discussed in Section~\ref{sec:method-objectives}: rapid motion and behavior transitions both fall under case (i) (pose alone is insufficient to characterize the moment), while occlusion-induced tracking failures fall under case (ii) (visual content cannot be inferred from pose alone). Appendix~\ref{app:pose-noise} tests case (iii) with synthetic pose errors, and Appendix~\ref{app:residual-recon} shows that occluded frames raise the residual without raising input-space reconstruction error.

\subsection{Residual Response to Pose Estimation Errors}
\label{app:pose-noise}

\begin{table}[!thp]
\centering
\caption{Share (\%) of corrupted CalMS21 slots selected at a 25\% residual budget, averaged over five seeds; without any effect the share would be 25\%. Burst is the number of consecutive frames over which a corruption is held; columns give the noise standard deviation as a multiple of $\sigma_0$.}
\label{tab:pose-noise}
\small
\begin{tabular}{l r ccccc}
\toprule
Corruption & Burst & $0.25\sigma_0$ & $0.5\sigma_0$ & $\sigma_0$ & $2\sigma_0$ & $4\sigma_0$ \\
\midrule
\emph{single} & 1 & 25.1 & 24.8 & 25.1 & 25.1 & 25.4 \\
\emph{single} & 8 & 25.5 & 25.6 & 25.0 & 25.8 & 29.8 \\
\emph{single} & 32 & 25.7 & 24.9 & 25.8 & 27.4 & 30.8 \\
\addlinespace[2pt]
\emph{translate} & 1 & 25.2 & 24.9 & 25.1 & 25.7 & 28.4 \\
\emph{translate} & 8 & 25.9 & 26.5 & 27.4 & 33.4 & 51.2 \\
\emph{translate} & 32 & 26.4 & 24.4 & 26.5 & 29.4 & 40.6 \\
\addlinespace[2pt]
\emph{independent} & 1 & 25.1 & 24.9 & 25.2 & 25.3 & 26.9 \\
\emph{independent} & 8 & 26.1 & 25.8 & 26.0 & 31.2 & 51.7 \\
\emph{independent} & 32 & 26.0 & 25.8 & 27.9 & 34.1 & 51.8 \\
\bottomrule
\end{tabular}
\end{table}

Case (iii) in Section~\ref{sec:method-objectives} concerns keypoints that are detected but misplaced, a failure mode that the keypoint-validity metric of Appendix~\ref{app:quantify} cannot detect. We test it on CalMS21, whose keypoints contain no missing detections and whose coordinates preserve absolute position, so that a whole-animal misdetection can be represented. Using the frozen pretrained motion encoder on the fine-tuning test set, we corrupt 5\% of the animal-frame slots and hold each corruption fixed over a burst of 1, 8, or 32 consecutive frames at the model frame rate, mimicking sustained tracking failures. Corruptions are applied in normalized coordinates and take three forms: displacing one random keypoint (\emph{single}), rigidly translating all keypoints of the animal, as when bedding is tracked instead of the mouse (\emph{translate}), and adding independent noise to every keypoint (\emph{independent}). The noise standard deviation is a multiple of $\sigma_0$, the standard deviation of keypoints around their per-frame centroid, which corresponds to about 65 and 36 pixels along the two image axes. We rank all slots by residual, select the top 25\%, and report the share of corrupted slots that are selected.

Isolated single-frame errors and small displacements barely change selection (Table~\ref{tab:pose-noise}). Sustained errors of larger magnitude are selected increasingly often: when all keypoints of an animal are displaced by $4\sigma_0$ over 8 frames, about half of the corrupted slots enter the top 25\% (51.2\% for \emph{translate} and 51.7\% for \emph{independent}), whereas displacing a single keypoint raises the share only to 29.8\%. The residual therefore responds to pose estimation errors in proportion to how strongly they distort the animal's pose.

\subsection{Residual versus Input-Space Reconstruction Error}
\label{app:residual-recon}

\begin{table}[!thp]
\centering
\caption{Quantization residual versus input-space reconstruction error on the fine-tuning test sets. Correlations are computed per animal and frame. The last four columns give the share (\%) of samples in each corner when both quantities are split at their quartiles; for Mouse Triplet, only tracked frames are included.}
\label{tab:residual-recon}
\resizebox{\linewidth}{!}{%
\begin{tabular}{l r cc cccc}
\toprule
Dataset & Samples & Pearson & Spearman & high $r_t$, low error & high $r_t$, high error & low $r_t$, low error & low $r_t$, high error \\
\midrule
Mouse Triplet & 23{,}994 & 0.39 & 0.65 & 0.7 & 13.6 & 15.5 & 0.8 \\
CalMS21 & 258{,}560 & 0.36 & 0.40 & 2.5 & 11.4 & 10.5 & 2.5 \\
Open Field & 227{,}136 & 0.49 & 0.61 & 1.4 & 14.7 & 12.8 & 0.6 \\
\bottomrule
\end{tabular}}
\end{table}

Section~\ref{sec:method-pretrain} distinguishes the residual from the input-space reconstruction error that prior work uses as an anomaly signal~\citep{gong2019memorizing, park2020learning}. To test whether the two carry different information, we measure how strongly they correlate and characterize the frames on which they disagree. We compare the residual $r_t$ (Eq.~\ref{eq:residual}) with the input-space reconstruction error of the RVQ decoder, defined as the per-frame mean squared error between the decoded $z_q$ and the input normalized coordinates, i.e., the quantity optimized by $\mathcal{L}_{\text{recon}}$. Both are computed per animal and frame on the fine-tuning test set with the frozen pretrained motion encoder, excluding padded frames. On Mouse Triplet, animals whose keypoints are missing, and whose inputs are therefore zero-imputed, are treated as occluded frames and analyzed separately from tracked frames.

The two quantities are positively but only moderately correlated on all three datasets (Table~\ref{tab:residual-recon}), as expected since both are computed from the same quantized code. The quartile split shows where they disagree. On Mouse Triplet, high residual with low reconstruction error is rare among tracked frames, yet 279 of the 3{,}810 occluded frames fall into this corner and make up 64.1\% of it. Frames in this corner have a median residual of 1.97 and a median reconstruction error of 0.0015, compared with 1.39 and 0.0034 over all tracked frames; occluded frames as a whole follow the same pattern, with medians of 1.84 and 0.0031. The decoder reproduces the pose it receives, so the reconstruction error stays low, whereas the residual flags that this pose does not fit the learned codebook. Rapid-motion frames, defined as the top quartile of inter-frame velocity excluding behavior transitions, instead raise both quantities: 31.1\% of them fall into the high-residual, high-error corner, compared with 13.6\% of all tracked frames. The two measures therefore agree while the pose remains usable and diverge once tracking fails.

\subsection{Social Attention Captures Interaction Structure Beyond Spatial Proximity}
\label{app:social_attention}

\begin{figure}[!thp]
  \centering
  \includegraphics[width=\linewidth]{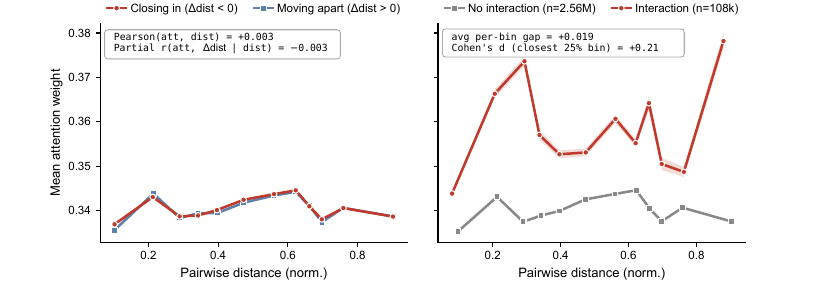}
\caption{\textbf{Cross-animal attention reflects interaction structure, not spatial proximity.} Head-averaged attention weight is plotted against pairwise distance (12 quantile bins) on (focal, partner, slot) records from Mouse Triplet. \textbf{Left panel: spatial geometry alone does not drive attention.} Records are split by approach direction (``closing in'' with $\Delta\text{dist} < 0$ vs.\ ``moving apart'' with $\Delta\text{dist} > 0$); the two lines are nearly flat and overlap, with Pearson $r$(attention, distance) $= +0.003$ and partial $r$(attention, $\Delta\text{dist}\mid$distance) $= -0.003$, both far below the conventional small-effect threshold $|r|\approx 0.1$. \textbf{Right panel: at the same distance, interaction frames still receive more attention than non-interaction frames.} Records are split by interaction status (whether the focal mouse has any behavior label active); at every matched-distance bin, interaction frames receive systematically higher attention, with a mean gap of $+0.019$ (Cohen's $d = +0.21$ at the closest $25\%$ bin.}
  \label{fig:social-attention}
\end{figure}

A natural concern with cross-animal social attention (Section~\ref{sec:contextual}) is that it might collapse into a distance kernel, simply attending to whichever partner is spatially closest regardless of behavioral interaction. To test this, we collect the social-attention weights from all fine-tune and pretrain windows and record one row per (focal, partner, slot) triplet using pretrained checkpoint. For each record we extract the head-averaged attention weight, the pairwise distance between the two animals' mean keypoints, the distance derivative $\Delta \text{dist}$ (negative = closing in), and a binary interaction label indicating whether the focal animal has any behavior class active at this slot. We bin all records into 12 quantile bins of distance and compare attention across two contrasts.

Figure~\ref{fig:social-attention}(left) splits records by approach direction. The two lines are nearly flat and lie almost on top of each other: the Pearson correlation between attention and distance is $r = +0.003$, and even when comparing pairs at similar distances, attention is essentially uncorrelated with whether they are approaching or receding (partial $r = -0.003$). Both are far below the conventional ``small effect'' threshold of $|r|\approx 0.1$, indicating that attention is driven neither by how close two animals are nor by whether they are approaching or receding. Figure~\ref{fig:social-attention}(right) instead splits records by interaction status. At every matched-distance bin, interaction-labeled pairs receive systematically higher attention, with a mean gap of $+0.019$ compared to no-interaction pairs. Then, interaction-labeled pairs also receive a standardized effect size of Cohen's $d = +0.21$ at the closest $25\%$ bin (Cohen's $d$ measures the gap in standard-deviation units, with $0.2$ conventionally a ``small'' but reliably detectable effect). Because the bins are matched on distance, this gap cannot be explained by spatial proximity: a chase and a passing-by at the same instantaneous distance do not produce the same attention pattern. The social-attention layer therefore captures behaviorally meaningful structure orthogonal to pure spatial geometry, validating the design choice of cross-animal attention over a hand-coded distance kernel.

\subsection{Why Not Cluster RVQ Codes Directly?}
\label{app:rvq_vs_dec}

\begin{figure}[!thp]
  \centering
  \includegraphics[width=\linewidth]{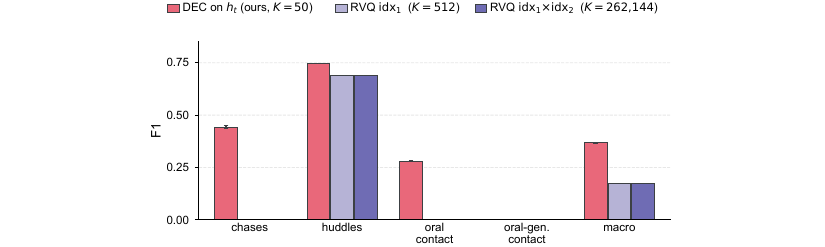}
\caption{\textbf{Clustering RVQ codebook indices directly fails to recover behavior categories.} Macro and per-class F1 on Mouse Triplet under the no-vision setting ($r=0\%$), comparing DEC on the learned fused representation $h_t$ (ours, $K=50$) against direct clustering on level-1 RVQ codes ($K=512$) and on the Cartesian product of both RVQ levels ($K=262{,}144$). All three pipelines use the same motion-only input and the same count-based majority-vote evaluation protocol. DEC reaches $2.1\times$ the macro F1 of either RVQ variant. Both RVQ variants collapse to the majority class \emph{huddles}, with $F1=0$ on the other three behaviors; enlarging the codebook from $512$ to $262{,}144$ does not help. DEC instead produces non-trivial $F1$ on three of four behaviors.}
  \label{fig:rvq-vs-dec}
\end{figure}

A natural alternative to DEC is to use the RVQ codebook indices themselves as cluster assignments, which would avoid the extra clustering step entirely. We test this on Mouse Triplet under the no-vision setting ($r=0\%$, motion-only representation) so that any performance difference is attributable to the clustering representation rather than to the mapping protocol or to visual side-channels. We compare three pipelines under the same count-based majority-vote evaluation: DEC on the fused representation $h_t$ ($K=50$, our default), level-1 RVQ codes ($K=512$), and the Cartesian product of both RVQ levels ($K=262{,}144$).

Direct RVQ clustering fails (Figure~\ref{fig:rvq-vs-dec}). Both RVQ variants collapse to the majority class \emph{huddles}, achieving a macro $F1$ of $0.172$, while DEC on $h_t$ reaches macro $F1 = 0.368$ on the same input under the same unsupervised protocol. Increasing the codebook size by $512\times$ ($K=512 \to 262{,}144$) yields no improvement, so the failure is not a matter of having too few clusters. Because only the clustering representation differs across the three pipelines, the bottleneck is the representation, not the mapping algorithm.

The reason is the difference in what each representation encodes. An RVQ code is a per-frame label assigned by a single codebook lookup on $z_e$, before the temporal and social attention modules are applied; it captures only the instantaneous kinematic state of one animal. The fused representation $h_t$, in contrast, is the output of the full motion encoder and carries temporal context (how this frame relates to nearby frames) and social context (how this animal relates to others). Behaviors such as \emph{chases}, \emph{oral contact}, and \emph{oral-genital contact} are defined precisely by these contextual relationships: a chase is a sustained pattern of one animal closely following another over time, and oral contact is defined by the relative configuration of two animals. RVQ codes do not see this structure, so frames from these three behaviors share overlapping single-frame kinematics with \emph{huddles} and end up assigned to codes that the majority-vote map labels as \emph{huddles}. DEC operating on $h_t$ has access to the contextual structure and can form clusters along behavior-relevant directions even without supervision. This justifies our choice to cluster on $h_t$ rather than to treat RVQ codes as off-the-shelf behavior cluster ids.

\subsection{Where Does Cross-Modal Pretraining Help?}
\label{app:vision_pretrain_disagreement}

\begin{figure}[!thp]
  \centering
  \includegraphics[width=0.75\linewidth]{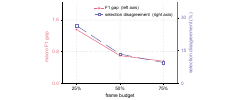}
\caption{\textbf{Selection disagreement and F1 gap shrink together as frame budget grows.} On Mouse Triplet supervised, we compare two pretrain regimes (with- vs.\ without-vision pretraining) at three intermediate budgets, holding the fine-tuning vision encoder, residual selection mechanism, and supervised loss fixed. The pink solid curve (left axis) is the F1 gap, with-vision minus without-vision in percentage points. The purple dashed curve (right axis) is selection disagreement, the fraction of frames selected by one pretrain but not the other (symmetric-difference rate between their top-budget frame sets) in fine-tuning test set. Both curves decrease monotonically and approximately in parallel as the budget grows.}
  \label{fig:vision-pretrain-disagreement}
\end{figure}

To understand \emph{where} cross-modal pretraining helps, we ask whether the with- and without-vision pretrains produce different residual rankings, and whether those ranking differences translate into the downstream F1 gap shown in Figure~\ref{fig:ablations}~(right). We hold every fine-tuning component fixed across the two regimes (same VideoPrism encoder, same residual selection, same supervised loss); the only manipulated variable is whether the cross-modal vision-prediction loss $\mathcal{L}_{\text{pred}}$ is active during pretraining. At each fine-tuning budget we compute two quantities. The first is the F1 gap, defined as (with-vision F1) $-$ (without-vision F1) in percentage points, which measures the downstream benefit of cross-modal pretraining. The second is selection disagreement, the fraction of frames selected by one pretrain but not the other at the same budget threshold (symmetric-difference rate), which is a purely ranking-level statistic computed from residual values alone, with no downstream loss involved.

The two curves decrease monotonically and approximately in parallel as the budget grows (Figure~\ref{fig:vision-pretrain-disagreement}). At $25\%$ budget the two pretrains disagree on $26.4\%$ of selected frames and the corresponding F1 gap is $1.36$ percentage points; at $75\%$ both quantities have shrunk to $9.5\%$ and $0.56$ points respectively. Because disagreement is a ranking-level statistic with no access to labels or task loss, its alignment with the task-level F1 gap is mechanism evidence rather than correlation: cross-modal pretraining does not merely shape the codebook in some abstract sense but changes which specific frames the residual flags as visually informative, and these changes translate into measurably different downstream classification. The alignment also explains why the F1 gap closes at high budgets: when most frames already receive vision, the small set of frames over which the two pretrains still disagree is too small to drive a large performance difference.

\subsection{Partial versus Full Vision Budget}
\label{app:budget-stats}

\begin{table}[!thp]
\centering
\caption{Paired comparison of the 75\% and 100\% budgets in the supervised setting. $\Delta$ is the mean difference in macro F1 (75\% minus 100\%) across seeds, where each run is summarized by the mean over its last three epochs; $p$-values are from a paired $t$-test.}
\label{tab:budget-stats}
\small
\begin{tabular}{l c c c c c}
\toprule
Dataset & Seeds & $\Delta$ & 95\% CI & $p$ & Seeds favoring 75\% \\
\midrule
Mouse Triplet & 10 & $+0.0097$ & $[+0.0068, +0.0127]$ & $<10^{-4}$ & 10/10 \\
Open Field & 10 & $+0.0149$ & $[-0.0188, +0.0487]$ & $0.34$ & 5/10 \\
CalMS21 & 5 & $-0.0156$ & $[-0.0196, -0.0115]$ & $<10^{-3}$ & 0/5 \\
\bottomrule
\end{tabular}
\end{table}

We test whether the difference between the 75\% and 100\% budgets is reliable by repeating supervised fine-tuning over multiple seeds (Table~\ref{tab:budget-stats}). For each seed, both budgets start from the same initialization, and the train, validation, and test splits are fixed, so the runs are paired. To avoid selecting epochs on the test set, each run is summarized by its mean macro F1 over the last three epochs. The 75\% budget is significantly better on Mouse Triplet, where it wins in every seed; on Open Field it is higher on average but not significantly; on CalMS21 the full budget is significantly better.

\begin{table}[!thp]
\centering
\caption{Per-class F1 on Mouse Triplet at the 75\% and 100\% budgets (mean over 10 seeds, last three epochs).}
\label{tab:budget-perclass}
\small
\begin{tabular}{l c c c}
\toprule
Behavior & 75\% & 100\% & $\Delta$ \\
\midrule
chases & 0.823 & 0.823 & $0.000$ \\
huddles & 0.945 & 0.946 & $-0.002$ \\
oral contact & 0.821 & 0.789 & $+0.032$ \\
oral-genital contact & 0.536 & 0.528 & $+0.008$ \\
\midrule
macro & 0.781 & 0.771 & $+0.010$ \\
\bottomrule
\end{tabular}
\end{table}

On Mouse Triplet, the advantage of the 75\% budget is concentrated in oral contact (Table~\ref{tab:budget-perclass}). At the full budget, oral-contact recall drops from 0.94 to 0.83 and false positives on the visually similar oral-genital contact rise by about 14\%, so oral-contact frames are increasingly labeled as oral-genital contact. Three analyses of the trained models locate the cause. First, the lost frames are not the low-residual frames to which the full budget adds vision: among frames that the 75\% models classify correctly but the 100\% models miss, only 2.5\% belong to this set, compared with 12.9\% of all oral-contact frames. Second, the information separating the two classes is preserved: a linear probe on the fused representations of high-residual oral-contact and oral-genital frames separates them equally well at both budgets (0.79). Third, the geometry of the representation shifts: the share of oral-contact frames closer to the oral-genital centroid than to their own rises from 5.0\% to 6.9\%, with no overlap between the ten models of each budget. Adding vision to low-residual frames thus perturbs the fused representation indirectly and moves a subset of oral-contact frames toward the neighboring class.

\begin{table}[!thp]
\centering
\caption{Fixed amount of vision on Mouse Triplet (supervised): 25\% of frames receive visual features, chosen by different rules. Macro F1 (mean $\pm$ std) follows Table~\ref{tab:main}, over three runs (five for the lowest-residual rule); $p$-values are from Welch's $t$-test against no vision.}
\label{tab:fixed-budget}
\small
\begin{tabular}{l c c c}
\toprule
Frames receiving vision & Macro F1 & $\Delta$ vs.\ no vision & $p$ \\
\midrule
None & 0.654\pgstd{0.005} & -- & -- \\
Uniform 25\% & 0.666\pgstd{0.006} & $+0.012$ & 0.050 \\
Highest-residual 25\% & 0.680\pgstd{0.002} & $+0.026$ & 0.008 \\
Lowest-residual 25\% & 0.635\pgstd{0.003} & $-0.019$ & 0.013 \\
\bottomrule
\end{tabular}
\end{table}

A partial budget could also help simply because less visual input must be fit from limited labels. To separate this effect from the choice of frames, we fix the amount of vision at 25\% and vary only which frames receive it (Table~\ref{tab:fixed-budget}). With the same amount of vision and the same labels, macro F1 ranges from $+0.026$ to $-0.019$ relative to no vision, and vision on the lowest-residual frames is worse than none. The outcome therefore depends on which frames are selected rather than on how much vision is used.

\subsection{Computational Cost}
\label{app:cost}

\begin{table}[!thp]
\centering
\caption{Cost of processing one 6.67\,s Mouse Triplet window (50 frames per mouse, three mice) on a single GPU. Frames is the number of frames encoded by the vision encoder. Retrieval decodes the selected frames from the compressed video; processing is the vision encoder forward pass; model is the fusion model forward pass. Macro F1 follows Table~\ref{tab:main}.}
\label{tab:cost}
\resizebox{\linewidth}{!}{%
\begin{tabular}{l r rrr r r c}
\toprule
Budget & Frames & Retrieval (ms) & Processing (ms) & Processing (TFLOPs) & Model (ms) & Total (ms) & Macro F1 \\
\midrule
25\% & 37.5 & 89 & 212 & 7.1 & 4 & 305 & 0.680 \\
50\% & 75 & 87 & 424 & 14.2 & 4 & 515 & 0.744 \\
75\% & 112.5 & 92 & 636 & 21.3 & 4 & 732 & 0.784 \\
100\% & 150 & 105 & 848 & 28.5 & 4 & 957 & 0.777 \\
\bottomrule
\end{tabular}}
\end{table}

\begin{table}[!thp]
\centering
\caption{Training and inference time. Pretraining uses two GPUs and 150 epochs; fine-tuning (50 epochs, training steps only) and inference (fusion model per window) use one GPU. None of these depends on the budget.}
\label{tab:traintime}
\small
\begin{tabular}{l c c c}
\toprule
Dataset & Pretraining (h) & Fine-tuning (min) & Inference (ms/window) \\
\midrule
Mouse Triplet & 8.3 & 6.9 & 4.3 \\
Open Field & 0.93 & 3.0 & 5.3 \\
CalMS21 & 0.21 & 4.6 & 3.4 \\
\bottomrule
\end{tabular}
\end{table}

Processing a window involves three stages: retrieving frames from the compressed video, encoding them with the vision encoder, and running the fusion model. Because frame selection depends only on pose, the first two stages apply only to the selected frames, and computing the residuals for selection takes a single pose-only pass over the dataset.

The spatial layers of VideoPrism process each frame independently, and its cost stays at 0.1897 TFLOPs per frame, varying by less than 0.05\% for inputs of 1 to 16 frames, so the encoder cost is linear in the number of encoded frames. Its latency is likewise linear, with a marginal cost of 5.65\,ms per frame, from which the processing times in Table~\ref{tab:cost} are computed. Retrieval is measured with the scattered frame access that residual selection produces and saves only sublinearly, because non-adjacent frames require seeking in the compressed video.

Processing dominates the total cost (70--89\%) and scales linearly with the budget (Table~\ref{tab:cost}). The 75\% budget reaches the best macro F1 at 76\% of the full-budget time, and the 50\% budget recovers about two thirds of the gain over no vision at 54\% of the time. A 6.67\,s window is processed 7$\times$ faster than real time at the full budget and 22$\times$ faster at 25\%. Training cost does not depend on the budget (Table~\ref{tab:traintime}): pretraining always uses visual features from all frames, and fine-tuning masks precomputed features according to the selection (Appendix~\ref{app:sup-details}).

\subsection{Vision Encoder Comparison}
\label{app:encoder-frames}

\begin{table}[!thp]
\centering
\caption{Macro F1 on Mouse Triplet (supervised) for each vision encoder and budget, with residual selection at intermediate budgets (values behind Figure~\ref{fig:encoder-analysis}). The best budget per encoder is in bold.}
\label{tab:encoder-budget}
\small
\begin{tabular}{l ccccc}
\toprule
Encoder & 0\% & 25\% & 50\% & 75\% & 100\% \\
\midrule
VideoPrism & 0.654 & 0.680 & 0.744 & \textbf{0.784} & 0.777 \\
VideoMAE(V2) & 0.627 & 0.629 & 0.718 & 0.745 & \textbf{0.768} \\
DINOv3 & 0.640 & 0.687 & 0.743 & \textbf{0.746} & 0.737 \\
CLIP & 0.618 & 0.665 & 0.698 & 0.719 & \textbf{0.738} \\
Qwen3.5 & 0.629 & 0.674 & 0.681 & \textbf{0.706} & 0.685 \\
\bottomrule
\end{tabular}
\end{table}

\begin{table}[!thp]
\centering
\caption{Frame-level linear probe on cached vision features (Mouse Triplet, single seed): test macro F1 when features are averaged over a centered window of $k$ frames, and the change from $k=1$ to $k=9$.}
\label{tab:encoder-probe}
\small
\begin{tabular}{l cccc}
\toprule
Encoder & $k=1$ & $k=5$ & $k=9$ & Change \\
\midrule
VideoPrism & 0.604 & 0.568 & 0.563 & $-0.042$ \\
VideoMAE(V2) & 0.487 & 0.475 & 0.468 & $-0.018$ \\
DINOv3 & 0.507 & 0.520 & 0.493 & $-0.013$ \\
CLIP & 0.501 & 0.483 & 0.539 & $+0.038$ \\
Qwen3.5 & 0.504 & 0.492 & 0.522 & $+0.018$ \\
\bottomrule
\end{tabular}
\end{table}

Table~\ref{tab:encoder-budget} lists the macro F1 behind Figure~\ref{fig:encoder-analysis}. A 50\% budget already recovers 65--97\% of each encoder's maximum gain over no vision.

To examine the CLIP exception in Section~\ref{sec:encoder-analysis}, we probe the cached per-frame features of each encoder. Features are averaged over the mice in each frame, reduced to 256 dimensions by PCA fitted on the training split, and averaged over a centered window of $k$ frames; a one-vs-rest logistic regression with balanced class weights is trained for each behavior and evaluated by macro F1 on the test split (Table~\ref{tab:encoder-probe}). CLIP gains the most from averaging neighboring frames ($+0.038$ from $k=1$ to $k=9$, compared with $+0.018$ for Qwen3.5 and negative changes for the other encoders), consistent with its full-vision baseline exceeding the 75\% budget. The probe uses a single seed, and the gain for CLIP is not monotonic in $k$.

\newpage

\input{checklist.tex}

\end{document}

%% file: checklist.tex
\section*{NeurIPS Paper Checklist}

\begin{enumerate}

\item {\bf Claims}
    \item[] Question: Do the main claims made in the abstract and introduction accurately reflect the paper's contributions and scope?
    \item[] Answer: \answerYes{} 
    \item[] Justification: The abstract and Section~\ref{sec:intro} state the contributions, which are supported by the experiments in Section~\ref{sec:experiment}.

\item {\bf Limitations}
    \item[] Question: Does the paper discuss the limitations of the work performed by the authors?
    \item[] Answer: \answerYes{} 
    \item[] Justification: Section~\ref{sec:conclusion} discusses the limitations: per-dataset training, visual features of all frames during pretraining, and a single budget ratio fixed per dataset.

\item {\bf Theory assumptions and proofs}
    \item[] Question: For each theoretical result, does the paper provide the full set of assumptions and a complete (and correct) proof?
    \item[] Answer: \answerNA{} 
    \item[] Justification: The paper does not include theoretical results.

    \item {\bf Experimental result reproducibility}
    \item[] Question: Does the paper fully disclose all the information needed to reproduce the main experimental results of the paper to the extent that it affects the main claims and/or conclusions of the paper (regardless of whether the code and data are provided or not)?
    \item[] Answer: \answerYes{} 
    \item[] Justification: Appendices~\ref{app:implementation} and~\ref{app:experiment} give the model, training, and evaluation details, and the code and pretrained weights are released (see the next question).

\item {\bf Open access to data and code}
    \item[] Question: Does the paper provide open access to the data and code, with sufficient instructions to faithfully reproduce the main experimental results, as described in supplemental material?
    \item[] Answer: \answerYes{} 
    \item[] Justification: Code, configurations, data splits, and pretrained weights are available at \url{https://github.com/BRAINML-GT/ResiFuse}, and the re-estimated Mouse Triplet pose at \url{https://huggingface.co/datasets/WeihanLikk/ResiFuse-MABe22}; the repository README gives the commands to reproduce the main results. Videos and labels come from the original releases of the datasets.

\item {\bf Experimental setting/details}
    \item[] Question: Does the paper specify all the training and test details (e.g., data splits, hyperparameters, how they were chosen, type of optimizer) necessary to understand the results?
    \item[] Answer: \answerYes{} 
    \item[] Justification: Appendices~\ref{app:implementation} and~\ref{app:experiment} give the data splits, hyperparameters, and optimizers.

\item {\bf Experiment statistical significance}
    \item[] Question: Does the paper report error bars suitably and correctly defined or other appropriate information about the statistical significance of the experiments?
    \item[] Answer: \answerYes{} 
    \item[] Justification: Table~\ref{tab:main} reports the mean and standard deviation over seeds, and Table~\ref{tab:budget-stats} gives paired comparisons with confidence intervals and $p$-values.

\item {\bf Experiments compute resources}
    \item[] Question: For each experiment, does the paper provide sufficient information on the computer resources (type of compute workers, memory, time of execution) needed to reproduce the experiments?
    \item[] Answer: \answerYes{} 
    \item[] Justification: Appendix~\ref{app:pretrain-details} lists the GPUs, and Appendix~\ref{app:cost} (Tables~\ref{tab:cost} and~\ref{tab:traintime}) gives the training and inference times.
    
\item {\bf Code of ethics}
    \item[] Question: Does the research conducted in the paper conform, in every respect, with the NeurIPS Code of Ethics \url{https://neurips.cc/public/EthicsGuidelines}?
    \item[] Answer: \answerYes{} 
    \item[] Justification: We have reviewed the NeurIPS Code of Ethics and confirm that this work conforms to it. The research uses three publicly available mouse behavior datasets (Open Field, CalMS21, Mouse Triplet) released by their respective authors for research purposes; no new animal experiments were conducted, and no human subjects or personally identifiable data are involved. We discuss the limitations of our approach in the main text.

\item {\bf Broader impacts}
    \item[] Question: Does the paper discuss both potential positive societal impacts and negative societal impacts of the work performed?
    \item[] Answer: \answerYes{} 
    \item[] Justification: Appendix~\ref{app:broader_impacts} discusses the potential positive and negative societal impacts.
    
\item {\bf Safeguards}
    \item[] Question: Does the paper describe safeguards that have been put in place for responsible release of data or models that have a high risk for misuse (e.g., pre-trained language models, image generators, or scraped datasets)?
    \item[] Answer: \answerNA{} 
    \item[] Justification: The released models and data describe laboratory mouse behavior and pose no high risk of misuse.

\item {\bf Licenses for existing assets}
    \item[] Question: Are the creators or original owners of assets (e.g., code, data, models), used in the paper, properly credited and are the license and terms of use explicitly mentioned and properly respected?
    \item[] Answer: \answerYes{} 
    \item[] Justification: The datasets (Open Field~\citep{sturman2020deep}, CalMS21~\citep{sun2021multi}, and MABe22~\citep{sun2023mabe22}) and pretrained models we use are cited where they are introduced and used under their terms; CalMS21 and MABe22 are distributed under CC BY-NC-SA, and our re-estimated Mouse Triplet pose is released under the same license.

\item {\bf New assets}
    \item[] Question: Are new assets introduced in the paper well documented and is the documentation provided alongside the assets?
    \item[] Answer: \answerYes{} 
    \item[] Justification: We release the code (MIT license), pretrained weights, and the re-estimated Mouse Triplet pose (CC BY-NC-SA 4.0); the repository README and the dataset card document their contents, formats, and usage.

\item {\bf Crowdsourcing and research with human subjects}
    \item[] Question: For crowdsourcing experiments and research with human subjects, does the paper include the full text of instructions given to participants and screenshots, if applicable, as well as details about compensation (if any)? 
    \item[] Answer: \answerNA{} 
    \item[] Justification: The paper does not involve crowdsourcing or research with human subjects.

\item {\bf Institutional review board (IRB) approvals or equivalent for research with human subjects}
    \item[] Question: Does the paper describe potential risks incurred by study participants, whether such risks were disclosed to the subjects, and whether Institutional Review Board (IRB) approvals (or an equivalent approval/review based on the requirements of your country or institution) were obtained?
    \item[] Answer: \answerNA{} 
    \item[] Justification: The paper does not involve crowdsourcing or research with human subjects.

\item {\bf Declaration of LLM usage}
    \item[] Question: Does the paper describe the usage of LLMs if it is an important, original, or non-standard component of the core methods in this research? Note that if the LLM is used only for writing, editing, or formatting purposes and does \emph{not} impact the core methodology, scientific rigor, or originality of the research, declaration is not required.
    \item[] Answer: \answerNA{} 
    \item[] Justification: LLMs are not part of the method; the vision tower of Qwen3.5 is only one of the vision encoders compared in Section~\ref{sec:encoder-analysis}.

\end{enumerate}